\documentclass[aoas]{imsart}

\RequirePackage{amsthm,amsmath,amsfonts,amssymb}
\RequirePackage[authoryear]{natbib}
\RequirePackage[colorlinks,citecolor=blue,urlcolor=blue]{hyperref}
\RequirePackage{graphicx}
\RequirePackage{bm}
\RequirePackage{mathrsfs}
\RequirePackage{multirow}
\RequirePackage{booktabs}
\usepackage{caption}
\usepackage{subcaption}
\usepackage[usenames,dvipsnames]{color}

\usepackage{tikz}
\usetikzlibrary{trees}

\newcommand{\bfx}{\bm{x}}

\newcommand{\bfalpha}{\bm{\alpha}}
\newcommand{\bfbeta}{\bm{\beta}}

\newcommand{\bftheta}{\bm{\theta}}

\newcommand{\bfS}{\bm{S}}
\newcommand{\bfT}{\bm{T}}

\newcommand{\bfX}{\bm{X}}
\newcommand{\bfY}{\bm{Y}}

\newcommand{\bbE}{\mathbb{E}}
\newcommand{\bbI}{\mathbb{I}}
\newcommand{\bbP}{\mathbb{P}}

\newcommand{\bbS}{\mathbb{S}}

\newcommand{\calD}{\mathcal{D}}

\newcommand{\calM}{\mathcal{M}}
\newcommand{\calN}{\mathcal{N}}
\newcommand{\calO}{\mathcal{O}}
\newcommand{\calQ}{\mathcal{Q}}

\newcommand{\calT}{\mathcal{T}}

\startlocaldefs
\theoremstyle{plain}
\newtheorem{axiom}{Axiom}

\newtheorem{assumption}[axiom]{Assumption}

\theoremstyle{remark}

\endlocaldefs

\begin{document}
	
	\begin{frontmatter}
		\title{A Bayesian Machine Learning Approach for Estimating Heterogeneous Survivor Causal Effects: Applications to a Critical Care Trial}
		\runtitle{Heterogeneous Survivor Causal Effects}
		
		\begin{aug}
			\author[A]{\fnms{Xinyuan} \snm{Chen}\ead[label=e1,mark]{xchen@math.msstate.edu}},
			\author[B]{\fnms{Michael O.} \snm{Harhay}\ead[label=e2,mark]{mharhay@pennmedicine.upenn.edu}},
			\author[C,D]{\fnms{Guangyu} \snm{Tong}\ead[label=e3,mark]{guangyu.tong@yale.edu}}
			\and
			\author[C,D]{\fnms{Fan} \snm{Li}\ead[label=e4,mark]{fan.f.li@yale.edu}}
			\address[A]{Department of Mathematics and Statistics,
				Mississippi State University,
				\printead{e1}}
			
			\address[B]{Department of Biostatistics, Epidemiology and Informatics,
				Perelman School of Medicine, University of Pennsylvania,
				\printead{e2}}
			
			\address[C]{Department of Biostatistics,
				Yale University School of Public Health,
				\printead{e3,e4}}
			
			\address[D]{Center for Methods in Implementation and Prevention Science, Yale University School of Public Health,
				\printead{e3,e4}}
		\end{aug}
		
		\begin{abstract}
			
			Assessing heterogeneity in the effects of treatments has become increasingly popular in the field of causal inference and carries important implications for clinical decision-making. While extensive literature exists for studying treatment effect heterogeneity when outcomes are fully observed, there has been limited development in tools for estimating heterogeneous causal effects when patient-centered outcomes are truncated by a terminal event, such as death. Due to mortality occurring during study follow-up, the outcomes of interest are unobservable, undefined, or not fully observed for many participants, in which case principal stratification is an appealing framework to draw valid causal conclusions. Motivated by the Acute Respiratory Distress Syndrome Network (ARDSNetwork) ARDS respiratory management (ARMA) trial, we developed a flexible Bayesian machine learning approach to estimate the average causal effect and heterogeneous causal effects among the always-survivors stratum when clinical outcomes are subject to truncation. We adopted Bayesian additive regression trees (BART) to flexibly specify separate mean models for the potential outcomes and latent stratum membership. In the analysis of the ARMA trial, we found that the low tidal volume treatment had an overall benefit for participants sustaining acute lung injuries on the outcome of time to returning home, but substantial heterogeneity in treatment effects among the always-survivors, driven most strongly by biologic sex and the alveolar-arterial oxygen gradient at baseline (a physiologic measure of lung function and source of hypoxemia). These findings illustrate how the proposed methodology could guide the prognostic enrichment of future trials in the field.

		\end{abstract}
		
		\begin{keyword}
			\kwd{acute lung injury}
			\kwd{Bayesian additive regression trees}
			\kwd{causal inference} 
			\kwd{heterogeneity of treatment effects} 
			\kwd{principal stratification}
			\kwd{truncation by death}
		\end{keyword}
		
	\end{frontmatter}
	
	\section{Introduction} \label{sec:intro}

	Personalized medicine, whereby healthcare is tailored for each individual patient, is the pursuit of contemporary clinical research and practice. For healthcare practitioners and clinicians, achieving this goal hinges upon the successful detection and an understanding of the heterogeneity in participants' response to treatment strategies based on their individual characteristics. Capturing factors prognostic of a stronger or weaker response to a trial intervention is especially important in critical care, where conditions such as cardiogenic shock, sepsis, and acute respiratory failure are defined by syndromic criteria such that individuals with the same condition can vary in their biologic and clinical presentation, and thus optimal treatment strategies can vary among clinical populations. 
	
	While examination of treatment effect heterogeneity for short-term mortality is difficult due to the small sample sizes common in critical care trials \citep{harhay2014outcomes}, this outcome is at least available for all individuals, and recent innovations in statistical learning increasingly permit such examinations \citep{Hill2011,Hahn2020}. In contrast, the estimation of average treatment effects and conditional average treatment effects for clinically important non-mortality outcomes, such as duration of organ support (e.g., ventilation) or need for intensive care unit or hospital-level care (i.e., length of stay) are more intractable because they are not fully observed, or more generally said to be `truncated' by the event of death. This is because some participants do not survive to the time point when the non-mortality outcome, such as quality of life, can be measured, or for duration-based outcomes such as length of stay, these outcomes are truncated by the inter-current event of mortality such that the actual time to hospital discharge cannot be assessed. As a result, for those who do not survive until the end of the study, their non-mortality outcome measure is ambiguous. Though not uncommon, the direct survivors-only analysis can produce selection biases because the truncation by death occurs post-randomization and is often informative \citep{Harhay2019}.

	Our motivating application is the Acute Respiratory Distress Syndrome Network (ARDSNET) ARMA trial, which was an individually-randomized clinical trial that compared respiratory management during mechanical ventilation with a lower tidal volume ventilator strategy (6 mL/kg) versus a higher tidal volume ventilator strategy (12 mL/kg) for participants suffering from acute lung injury \citep{brower2000acute}. The first primary outcome of the ARMA trial was death before a participant was discharged home and was breathing without assistance, and, the second primary outcome was the number of days without ventilator use from day 1 to day 28. As interest in critical care is increasingly focused on longer-term and patient-centered outcomes, we focus our analysis on a slightly longer time horizon (but a highly correlated measure to the second primary outcome) by using the outcome of days to returning home (DTRH). As is the case in other critical care intervention studies, a substantial proportion of the randomized trial participants (34.3\%) died before being discharged from the hospital, leading to undefined DTRH outcomes in a third of the trial sample.
	
	In the ARMA trial, one of the few critical care trials that successfully identified a statistically significant treatment effect in the past three decades \citep{Tonelli2014,matthay2017clinical}, the survival status of participants is observed post-treatment assignment and is regarded as an intermediate variable. Under the potential outcome framework, \citet{Frangakis2002biometrics} proposed the principal stratification approach as a framework to define causal effects in the presence of an intermediate variable. In the context of the ARMA trial, the joint potential values of the survival status allow us to classify participants into distinct strata, and the potential outcomes of the non-mortality endpoint are only well-defined among the always-survivors (those who are likely in healthier or more treatment-responsive conditions at the time of randomization). Therefore the survivor average causal effect (SACE) can be considered as an interpretable principal causal effect for non-mortality outcomes \citep{ZhangRubin2003}. 

The existing literature for SACE can be largely categorized into two streams. The first stream involves deriving nonparametric large-sample bounds to interval identify the SACE under minimal assumptions, e.g., \citet{ZhangRubin2003}, \citet{imai2008sharp}, \citet{Ding2011}, \citet{long2013sharpening}, and \citet{Yang2016}. However, these bounds are often too wide to be informative for real (i.e., clinical or policy) applications. Beyond interval identification, the second stream of literature invokes additional structural and parametric modeling assumptions to identify SACE, e.g., \citet{Hayden2005}, \citet{Egleston2006}, \citet{Zhang2009jasa}, \citet{Chiba2011}, \citet{Frumento2012}, \citet{wang2017identification} and \citet{Bia2021}. While convenient to implement, fully parametric modeling necessitates restrictive assumptions that are often challenging to verify. In addition, the bulk of this literature has focused on the average causal effect among the always-survivors, and has not branched into understanding how the always-survivors may be deferentially affected by treatment due to their individual characteristics.

In this article, we address the goal of estimating the heterogeneous treatment effects among the always-survivors stratum in the ARMA tidal volumes trial using the patient-centered and health-systems relevant DTRH outcome that was informatively truncated by in-hospital death. The target estimand for our new approach is the conditional survivor average causal effect (CSACE), which is defined as the average causal effect for an always-survivor with certain baseline characteristics. Proceeding under the Bayesian principal stratification framework, we relax the parametric modeling assumptions by leveraging the Bayesian additive regression trees (BART) ensemble algorithm \citep{Chipman2010} for estimating both the stratum membership model as well as the stratum-specific potential outcome models. While several Bayesian nonparametric prior models are successfully adapted for the purpose of causal inference with and without an intermediate variable---e.g., Dirichlet process mixture models \citep{kim2017framework,kim2019bayesian} and dependent Dirichlet process-Gaussian process prior models \citep{xu2016bayesian,roy2017bayesian,Xu2022}, the BART prior model has gained substantial traction for causal inference due to its computational efficiency and flexibility in modeling complex nonlinear interactions with minimum tuning; see, for example, a comprehensive tutorial by \citet{tan2019bayesian}, and empirical evidence supporting the use of the BART approach for estimating heterogeneous causal effects in different contexts \citep{Henderson2018,wendling2018comparing,dorie2019automated,Hahn2020,Hu2021,Bargagli2022}. We therefore propose to integrate the BART priors into the mixture model framework as a computationally convenient and yet effective approach for principal stratification analysis, and use the proposed approach to reanalyze a high-profile critical care trial---the ARMA trial---to quantify treatment effect heterogeneity and identify key effect moderators among the always-survivors in a data-driven fashion. A unique feature of the ARMA trial is that there are baseline covariates that are predictive of the survival status and hence the principal strata membership; this important feature provides a strong basis for investigating treatment effect heterogeneity among the always-survivors.

The remainder of this article is organized as follows. Section \ref{sec:not-asp} provides a concise overview of the principal stratification framework. Section \ref{sec:inf} introduces the Bayesian machine learning approach for principal stratification analyses and describes the details of drawing posterior samples for estimation and inference. Section \ref{sec:app} provides a reanalysis of the ARMA trial using the proposed Bayesian machine learning method and identifies key effect moderators. Section \ref{sec:con} offers concluding remarks and discusses future extensions. 

\section{Notation and set up} \label{sec:not-asp}

We consider a two-arm randomized trial with $N$ participants in the setting of non-mortality outcome truncated by death. Let $T_i$ represent the binary treatment for participant $i$, where $T_i=1$ if participant $i$ is randomized to treatment and $T_i=0$ otherwise. Under the potential outcomes framework, we let $Y_i(t)$ represent the non-mortality outcome that would have been observed under treatment assignment $T_i=t\in\{0,1\}$, and $\{Y_i(1),Y_i(0)\}$ be a pair of potential outcomes for each participant corresponding to the treatment and control conditions. We further define $D_i(t)$ as the potential survival status of participant $i$ at the time that the measurement of the non-mortality outcome (e.g., quality-of-life or DTRH) was taken, with 0 indicating death and 1 indicating being alive. For example, in the analysis of the ARMA trial, we define the survival status at 180 days, which is considered as the maximum DTRH. Similarly, $\{D_i(1),D_i(0)\}$ are a pair of potential survival statuses. In what ensues, we use $D_i$ and $Y_i$ to denote, respectively, the observed survival status and observed non-mortality outcome for participant $i$. Of note, an alternative set up is to view both DTRH and time-to-death as two time-to-event outcomes under the semi-competing risks framework, and consider potential survival status as a function of follow-up time \citep{comment2019survivor,Xu2022,Nevo2021}. Although we primarily focus on defining the potential survival status at a specific time point due to clinical relevance and to simplify the exploration of heterogeneity of treatment effect on the non-mortality outcome, we discuss the applicability of the semi-competing risks framework to the ARMA trial at the end of Section \ref{sec:con}.

We first make the Stable Unit Treatment Value Assumption (SUTVA). The SUTVA implies that there is one version of the treatment and that there is no interference between participants so that each participant's outcome only depends on the participant's own treatment. Under the SUTVA, $D_i=T_iD_i(1)+(1-T_i)D_i(0)$, and $Y_i=T_iY_i(1)+(1-T_i)Y_i(0)$ for those who survived until the time that the non-mortality outcome is measured. The non-mortality outcome for those who did not survive ($D_i=0$) is undefined, and we supplementarily augment the definition of outcome such that $Y_i=*$ \citep{Zhang2009jasa}. Using the principal stratification framework \citep{Frangakis2002biometrics}, each participant can be classified into one of the distinct principal strata according to the joint values of the potential survival status. Specifically, we have the following four possible stratum memberships:
\begin{itemize}
	\item[(a)] $S_i=11$, $\{i|D_i(1)=1, D_i(0)=1\}$, always-survivors: participants who would survive to the time of outcome measurement under either treatment status;
	\item[(b)] $S_i=10$, $\{i|D_i(1)=1, D_i(0)=0\}$, protected: participants who would survive to the time of outcome measurement under treatment but would die before then under control;
	\item[(c)] $S_i=01$, $\{i|D_i(1)=0, D_i(0)=1\}$, harmed: participants who would die before the time of outcome measurement under treatment but would survive under control;
	\item[(d)] $S_i=00$, $\{i|D_i(1)=0, D_i(0)=0\}$, never-survivors: participants who would die before the time of outcome measurement under either treatment status.
\end{itemize}
Since the pair of non-mortality potential outcomes is only well-defined among the always-survivors, a common causal estimand of interest is the SACE, defined as
\begin{equation*}
	\Delta_{SACE}=\bbE[Y_i(1)-Y_i(0)|S_i=11].
\end{equation*}
This principal causal effect is derived by averaging the individual potential outcomes contrasts over the population of always-survivors, and serves as the basis for concluding effectiveness regarding the treatment without ambiguity in defining the potential outcomes. Assuming $\bfX_i=\bfx$ is the baseline characteristics of individual $i$, we are additionally interested in the CSACE, defined as
\begin{equation}\label{eq:CSACE}
	\Delta_{CSACE}(\bfx)=\bbE[Y_i(1)-Y_i(0)|\bfX_i=\bfx,S_i=11],
\end{equation}
which quantifies the conditional treatment effect given certain baseline characteristics of an always-survivor who would live to the time of outcome measurement regardless of treatment assignment. Variations in $\Delta_{CSACE}(\bfx)$ measure the degree of treatment effect heterogeneity among the always-survivors, and may provide useful evidence for tailoring treatment rules for future participants. \citet{Deng2021} discussed identification strategies for CSACE under truncation by death, but under more restrictive conditions such as principal ignorability \citep{ding2017principal}. In this article, we provide an estimation approach that does not invoke principal ignorability, and only requires the following two standard structural assumptions. 

\begin{assumption}\label{asp2} (Randomization). The assignment $T_i$ is independent of all potential outcomes $\{D_i(1),D_i(0),Y_i(1),Y_i(0)\}$, given baseline characteristics $\bfX_i$.
\end{assumption}

\begin{assumption}\label{asp3} (Monotonicity). $\bbP(D_i(1)\geq D_i(0)|\bfX_i=\bfx)=1$, $\forall~\bfx\in\mathcal{X}$, where $\mathcal{X}$ is the support of $\bfX$. 
\end{assumption}

Assumption \ref{asp2} is essentially an ignorability assumption and holds by design in a randomized trial. However, it is more general and can be satisfied in stratified randomized studies as well as observational studies as long as $\bm{X}_i$ captures a sufficient set of control variables. Assumption \ref{asp3} states that the treatment does not lead to poor survival, and rules out the harmed stratum. This assumption is often considered plausible in studies where a treatment is designed to improve the general well-being of participants, as in our motivating application. Under Assumption \ref{asp3}, trial participants belong to one of the three strata of always-survivors, protected, or never-survivors, and depending on the observed treatment status, only a fraction of participants have unobserved stratum membership. In other words, survivors in the treatment arm can be either always-survivors or protected; non-survivors in the control arm can be either never-survivors or protected. Assumption \ref{asp3} may be violated when, for example, in a comparative effectiveness trial where two active treatments with unknown relative benefits are studied. In that case, it is of interest to extend our approach along the lines of \citet{Zhang2009jasa} by incorporating the harmed stratum, at the expense of reduced precision and algorithm stability. We return to a discussion of this point in Section \ref{sec:con}. 


\section{A Bayesian machine learning approach for estimating CSACE} \label{sec:inf}

\subsection{Bayesian principal stratification}

We consider the Bayesian principal stratification framework \citep{hirano2000assessing,Mattei2007,Mattei2013}, in which one is required to specify two sets of models: the distribution of potential outcomes $Y(0)$ and $Y(1)$ conditional on the principal strata and covariates (the $Y$-model), and the distribution of principal strata conditional on the covariates (the $S$-model). Let $\bftheta$ generically denote the global parameters, and for participant $i$, we use $\bfX_{Y,i}$ and $\bfX_{S,i}$ to denote respective vectors of covariates for the $Y$- and $S$-model, with $\bfX_i=(\bfX_{Y,i}',\bfX_{S,i}')'$. According to their treatment assignments and survival status at the time of final outcome measurement, we can reclassify each participant into the following categories:
\begin{itemize}
	\item[(a)] $\calO(1,1)=\{i|T_i=1,D_i=1\}$, participants assigned to the treatment arm and survived;
	\item[(b)] $\calO(1,0)=\{i|T_i=1,D_i=0\}$, participants assigned to the treatment arm and died;
	\item[(c)] $\calO(0,1)=\{i|T_i=0,D_i=1\}$, participants assigned to the control arm and survived;
	\item[(d)] $\calO(0,0)=\{i|T_i=0,D_i=0\}$, participants assigned to the control arm and died.
\end{itemize}
Stratum memberships for participants in $\calO(1,0)$ and $\calO(0,1)$ are then fully inferred under the monotonicity assumption, which are denoted by $S_i^{obs}$. We use $\bfS^{obs}$ to denote the collection of $S_i^{obs}$'s. On the other hand, for participants in $\calO(1,1)$ and $\calO(0,0)$, their stratum memberships cannot be determined directly, and are thus labeled as $S_i^{mis}$. Denote $\pi_{s,i}=\bbP(S_i=s|\bfX_{S,i},\bftheta)$ and $f_{st,i}=\bbP(Y_i(t)|S_i=s,\bfX_{Y,i},\bftheta)$, for $s=00,10,11$ and $t=0,1$, and assume a prior distribution $\bbP(\bftheta)$ for the parameters $\bftheta$. The posterior distribution of $\bftheta$ can be generically written as 
\begin{equation} \label{eq:joint-post}
	\begin{aligned}
		\bbP(\bftheta|\bfY,\bfS^{obs},\bfT,\bfX) &\propto \bbP(\bftheta) \times \prod_{i\in\calO(1,1)}\left(\pi_{11,i}f_{111,i}+\pi_{10,i}f_{101,i}\right)\times\prod_{i\in\calO(1,0)}\pi_{00,i} \\
		&\quad\times\prod_{i\in\calO(0,1)}\pi_{11,i}f_{110,i}\times\prod_{i\in\calO(0,0)}\left(\pi_{10,i}+\pi_{00,i}\right).
	\end{aligned}  
\end{equation}

\subsection{Model specification}\label{sec:parametric}

Posterior inference on $\bftheta$ from \eqref{eq:joint-post} is achieved using data augmentation to impute missing stratum membership $S_i^{mis}$, which can be performed via a nested Probit modeling approach. We introduce two additional latent variables $Z$ and $W$ to be augmented for each participant, where 
\begin{equation} \label{eq:aug-zw}
	\begin{aligned}
		\{Z_i|m_Z(\bullet),\bfX_{S,i}\} &\sim \calN\left(m_Z(\bfX_{Z,i}),1\right), ~ \text{and} ~ \left\{\begin{array}{lc}
			S_i=00, & \text{if}~Z_i>0 \\
			S_i=10~\text{or}~11, & \text{if}~Z_i\leq0
		\end{array}\right. \\
		\{W_i|m_W(\bullet),\bfX_{S,i}\} &\sim \calN\left(m_W(\bfX_{W,i}),1\right), ~ \text{and} ~ \left\{\begin{array}{cc}
			S_i=10, & \text{if}~W_i>0 \\
			S_i=11, & \text{if}~W_i\leq0
		\end{array}\right.
	\end{aligned}.
\end{equation}
Here, $m_Z(\bullet)$ and $m_W(\bullet)$ are conditional mean functions for $Z_i$ and $W_i$ that can be fully specified by corresponding parameters, and $\bfX_{Z,i}$ and $\bfX_{W,i}$ are vectors of covariates that are subsets of $\bfX_{S,i}$ with possible overlapping elements. Based on \eqref{eq:aug-zw}, the conditional probability of stratum membership for each participant can be expressed as
\begin{align*}
	\bbP\left(S_i=00|m_Z(\bullet),\bfX_{S,i}\right) &= 
	1-\Phi\left(m_Z(\bfX_{Z,i})\right), \\
	\bbP\left(S_i=10|m_Z(\bullet),m_W(\bullet),\bfX_{S,i}\right) 
	&= \Phi\left(m_Z(\bfX_{Z,i})\right)\left\{1-\Phi\left(m_W(\bfX_{W,i})\right)\right\}, \\
	\bbP\left(S_i=11|m_Z(\bullet),m_W(\bullet),\bfX_{S,i}\right) 
	&= \Phi\left(m_Z(\bfX_{Z,i})\right)\Phi\left(m_W(\bfX_{W,i})\right),
\end{align*}
where $\Phi(\bullet)$ is the cumulative distribution function of a standard normal random variable. Connecting with the notation in \eqref{eq:joint-post}, we have $\pi_{00,i}=1-\Phi\left(m_Z(\bfX_{Z,i})\right)$, $\pi_{10,i}=\Phi\left(m_Z(\bfX_{Z,i})\right)\left\{1-\Phi\left(m_W(\bfX_{W,i})\right)\right\}$, and $\pi_{11,i}=\Phi\left(m_Z(\bfX_{Z,i})\right)\Phi\left(m_W(\bfX_{W,i})\right)$. 

For the $Y$-models, we specify the three sets of potential outcome models as
\begin{align*}
	\{Y_i(t)|S_i=s,m_{st}(\bullet),\bfX_{Y,i}\}\sim\calN\left(m_{st}(\bfX_{st,i}),\sigma_{st}^2\right),
\end{align*}
where $t=0,1$ for $s=11$, and $t=1$ for $s=10$; $m_{st}(\bullet)$ are conditional mean functions for $Y_i(t)$ with $\bfX_{st,i}$ being vectors of covariates that are subsets of $\bfX_{Y,i}$ with possible overlapping elements, and $\sigma_{st}^2$ is the variance parameter that depends on the principal strata and the treatment status. Similar to the conditional mean functions in the $S$-model, $m_{st}(\bullet)$ are also fully specified by corresponding parameters. To summarize, we have the following,
\begin{equation} \label{eq:outcome}
	\begin{aligned}
		& \{Y_i(1)|S_i=11,m_{111}(\bullet),\bfX_{Y,i}\} \sim \calN\left(m_{111}(\bfX_{111,i}),\sigma_{111}^2\right) \\
		& \{Y_i(0)|S_i=11,m_{110}(\bullet),\bfX_{Y,i}\} \sim \calN\left(m_{110}(\bfX_{110,i}),\sigma_{110}^2\right)\\
		& \{Y_i(1)|S_i=10,m_{101}(\bullet),\bfX_{Y,i}\} \sim \calN\left(m_{101}(\bfX_{101,i}),\sigma_{101}^2\right)
	\end{aligned}.
\end{equation}
Based on the specification in \eqref{eq:outcome}, the SACE can be estimated as
\begin{equation} \label{eq:sace-bayes}
	\widehat{\Delta}_{SACE}=\iint\left[\int_{\mathcal{X}}\left\{m_{111}(\bfX_{111})-m_{110}(\bfX_{110})\right\}f(\bfX|S=11)\mu(d\bfX)\right]p_m p_{S},
\end{equation}
where the outer double integration is taken with respect to $p_m$---the posterior distributions of parameters in $m_{111}(\bullet)$ and $m_{110}(\bullet)$, and $p_S$---the posterior distribution of the principal strata membership. In addition, the CSACE evaluated at $\bfX=\bfx$ can be estimated by
\begin{equation} \label{eq:isce-bayes}
	\widehat{\Delta}_{CSACE}(\bfx)
	=\int\left\{m_{111}(\bfx)-m_{110}(\bfx)\right\}p_m.
\end{equation}
Implicitly in the notation of \eqref{eq:sace-bayes} and \eqref{eq:isce-bayes}, we consider super-population inference (rather than finite-sample inference), where the causal estimand is represented by model parameters governing the joint distribution of potential outcomes. Under this framework, the posterior distribution of $m_{111}(\bullet)$ and $m_{110}(\bullet)$ does not involve the correlation parameter between $Y_i(1)$ and $Y_i(0)$ (as the observed data likelihood is free of this correlation parameter), and therefore it is sufficient to specify the marginal distributions of potential outcomes; also see Section 3 of \citet{Ding2018} for a detailed discussion on this issue. In Web Appendix A4, we also conduct an additional analysis of the ARMA trial to infer the finite-sample SACE estimand, varying the correlation between $Y_i(1)$ and $Y_i(0)$ as a sensitivity parameter. The results are almost identical to the super-population analysis.

According to \eqref{eq:sace-bayes} and \eqref{eq:isce-bayes}, a central task in the estimation of the SACE and CSACE is to specify the conditional mean functions in the models. Typically, we assume that the parameters in these models are \emph{a priori} independent and proceed with conjugate diffuse priors. For example, a straightforward specification for the conditional mean functions can be achieved via parametric linear models such that $m_Z(\bfX_{Z,i})=\bfX_{Z,i}'\bfbeta_Z$, $m_{W}(\bfX_{W,i})\allowbreak=\bfX_{W,i}'\bfbeta_W$, and $m_{st}(\bfX_{st,i})=\bfX_{st,i}'\bfalpha_{st}$. Then a closed-form Gibbs sampler can be derived with multivariate Gaussian prior assumed for linear coefficients, $\bfbeta_Z$, $\bfbeta_W$, and $\bfalpha_{st}$. A detailed derivation of this Gibbs sampler is provided in Web Appendix A1. This fully parametric specification, however, can result in potential biases for estimating the SACE and CSACE when the true mean functions are nonlinear and with possibly unknown functional forms. An illustration of the bias resulting from model misspecification is provided in Web Appendix A2 using simulated data sets.

\subsection{Integrating Bayesian additive regression trees into principal stratification}

To address the potential limitations of fully parametric models, we propose to use the Bayesian additive regression trees (BART) to estimate the mean functions nonparametrically. BART is an ensemble method in which the mean function of a regression is approximated by the sum of individual trees, with prior distributions imposed to regularize the fit by keeping the individual tree effects relatively small \citep{Chipman2010}. Specifically, let $\calT$ denote a binary tree consisting of a set of interior node decision rules and a set of terminal nodes, and let $\calM =\{\mu_1,\mu_2,\ldots,\mu_b\}$ denote a set of parameter values associated with each of the $b$ terminal nodes of $\calT$. The BART formulation of the mean function relies on a collection of $J$ binary trees $\{\calT_1,\ldots,\calT_J\}$ and their respectively associated set of terminal node values $\{\calM_1,\ldots,\calM_J\}$ for each binary tree, where $\calM_j =\{\mu_{j1},\mu_{j2},\ldots,\mu_{jb_j}\}$. Each tree $\calT_j$ consists of a sequence of decision rules through which any covariate vector can be assigned to one terminal node of $\calT_j$ by following the decision rules prescribed at each of the interior nodes. The decision rules at the interior nodes of $\calT_j$ are of the form $\{X_k\leq c\}$ versus $\{X_k>c\}$, where $X_k$ denotes the $k$th element of $\bfX$. A covariate $\bfX$ that corresponds to the $l$th terminal node of $\calT_j$ is assigned the value $\mu_{jl}$ and $h(\bfX;\calT_j,\calM_j)$ is used to denote the function returning $\mu_{jl} \in \calM_j$ whenever $\bfX$ is assigned to the $l$th terminal node of $\calT_j$. The mean function of a generic regression model, $m(\bfX)$, can thus be represented as a sum of individual trees
\begin{align*}
	m(\bfX)=\sum_{j=1}^J h(\bfX;\calT_j,\calM_j).
\end{align*}

Under the BART formulation, the trees $\calT_j$ and node values $\calM_j$ can be thought of as model parameters. The prior distribution on these parameters induces a prior on $h(\bfX;\calT_j,\calM_j)$ and hence induces a prior on the mean function $m(\bfX)$. To proceed, one needs to specify the following to complete the description of the prior on $(\calT_1,\calM_1),\ldots,(\calT_J,\calM_J)$: (i) the distribution on the choice of splitting variable at each internal node; (ii) the distribution of the splitting value $c$ used at each internal node; (iii) the probability that a node at a given node-depth $\delta$ splits, which is assumed to be equal to $\tau(1+\delta)^{-\gamma}$; and (iv) the distribution of the terminal node values $\mu_{jl}$. Regarding (i)-(iii), we defer to defaults suggested in \citet{Chipman2010}, where, for (i), the splitting variable is chosen uniformly from the set of available splitting variables at each interior node; for (ii), a uniform prior on the discrete set of available splitting values is adopted; for (iii), the depth-related hyperparameters are chosen as $\tau=0.95$ and $\gamma=2$. For (iv), the distribution of the terminal node values $\mu_{jl}$ is assumed to be $\mu_{jl}\sim \calN\{0, (4w^2J)^{-1}\}$, where $w$ and $J$ are determined via cross-validation as we further elaborate in Section \ref{sec:app}. To denote the distribution on the regression function $m(\bullet)$ induced by the prior distribution on $\calT_j$ and $\calM_j$ with parameter values $(\tau,\gamma,w)$ and $J$ total trees, we use the notation $m(\bullet)\sim \mathrm{BART}(\tau,\gamma,w,J)$. Using BART, the mean functions under the Bayesian principal stratification framework can be expressed as
\begin{equation}
	\begin{aligned}
		m_Z(\bfX_{Z,i}) &= \sum_{j=1}^{J_Z} h_Z(\bfX_{Z,i};\calT_{Z,j},\calM_{Z,j}),\\ m_W(\bfX_{W,i}) &= \sum_{j=1}^{J_W} h_W(\bfX_{W,i};\calT_{W,j},\calM_{W,j}),\\
		m_{st}(\bfX_{st,i}) &= \sum_{j=1}^{J_{st}} h_{st}(\bfX_{st,i};\calT_{st,j},\calM_{st,j}),~~~st\in\{111,110,101\}
	\end{aligned}
\end{equation}
each with the common prior distribution $\mathrm{BART}(\tau,\gamma,w,J)$ that is assumed to be \emph{a priori} independent of each other; here $m_Z(\bullet)$, $m_W(\bullet)$ stand for the mean functions of the stratum membership model, $m_{st}(\bullet)$ stands for the mean function of the potential outcome model. Essentially, our semiparametric model is a mixture of BART, with the mixture weights represented by a nested Probit BART model. 

It is worth mentioning that with only Assumptions \ref{asp2} and \ref{asp3} alone, the SACE and CSACE estimands are only partially or set identified \citep{Kadane1975}; see, for example, the large sample bounds developed in Table 6 of \citet{ZhangRubin2003} and discussions of a similar issue arising from the noncompliance context in the Appendix of \citet{Hahn2016}. Resembling the approach of \citet{hirano2000assessing} for studying noncompliance, the parametric mixture approach in Section \ref{sec:parametric} resolves the partial identification issue through prior probability modeling, so that the inferential procedure is assisted by a Gaussian mixture model. The proposed approach considers the same prior probability modeling idea to resolve partial identification issue, but incorporates the BART priors for mean functions (still within a Gaussian mixture model) to more flexibly study effect moderation. Although not pursued in the current work, an alternative and powerful approach to address partial identification can follow \citet{Hahn2016} to transparently separate identified and unidentified model components with careful prior specifications for each component. Finally, we note that the proposed approach for imposing monotonicity differs from the recent work of \citet{Papakostas2021}. Whereas \citet{Papakostas2021} adopted BART priors for components of a compositional representation to enforce a stochastic monotonicity constraint (in the absence of intermediate variables), we adopted BART priors for mean functions in a Gaussian mixture model where the mixture itself already incorporates a structural monotonicity constraint. Finally, we emphasize that our mixture modeling approach is particularly suitable for studying CSACE when there exist baseline covariates that are predictive of the principal strata membership. This is the case in the ARMA trial and allows us to meaningfully describe the subset of always-survivors, as we exemplify in Section \ref{sec:match}.

\subsection{Posterior computation}\label{sec:post}

For posterior inference, we develop a Gibbs sampling procedure based upon the original Metropolis-within-Gibbs sampler proposed in \citet{Chipman2010}, which works by sequentially updating each tree while holding all other $J-1$ trees fixed. As a result, each iteration of the Gibbs sampler consists of $2J+1$ steps where the first $2J$ steps involve updating either one of the trees $\calT_j$ or terminal node parameters $\calM_j$ and the last step involves updating the residual variance parameter. 
In each iteration, we first update values of $Y$-model parameters by sampling from their respective full conditional posterior distributions and follow up by updating $S$-model parameters via sampling from related full conditional posterior distributions. Unobserved stratum memberships and additional latent variables, $Z$ and $W$, in the $S$-model are handled through additional data augmentation steps \citep{albert1993bayesian}. In essence, each component BART model can be updated separately because (i) an independent BART prior is assumed for each component model; (ii) the full conditional distributions for each component $Y$-model only depend on the observed potential outcomes, augmented stratum memberships, covariate vector, and residual variances, as well as relevant prior distributions; and (iii) the full conditional distributions for each component $S$-model only depend on the augmented latent variable, covariate vector, and relevant prior distributions. A detailed outline of the sampling procedure is as follows:
\begin{enumerate}
\item[1.] Update trees $\{\calT_{st,1},\dots,\calT_{st,J_{st}}\}$ and node parameters $\{\calM_{st,1},\dots,\calM_{st,J_{st}}\}$ in the $Y$-model via the Bayesian backfitting approach of \citet{Chipman2010}, using $Y_i(T_i)$ with $T_i=t$ in strata $S_i=s$ as responses. Full conditional distributions of trees and node parameters in the $Y$-model depend on the observed potential outcomes, $Y_i(T_i)$, latent stratum memberships, $S_i$, covariate vector $\bfX_{st,i}$, residual variances, $\sigma_{st}^2$, and prior distributions of trees and node parameters. Update values of the mean functions, $m_{st}(\bfX_{st,i})$ for $i=1,\dots,n$, using the updated $\{\calT_{st,1},\dots,\calT_{st,J_{st}}\}$ and $\{\calM_{st,1},\dots,\calM_{st,J_{st}}\}$, where $t=0,1$ for $s=11$, and $t=1$ for $s=10$. These mean function values are needed for the update of residual variances.
\item[2.] Update variance parameters in each $Y$-model. Assuming a conjugate inverse Gamma prior distribution $IG(a_0,b_0)$ for residual variance, $\sigma_{st}^2$ with $t=0,1$ for $s=11$ and $t=1$ for $s=10$, we update $\sigma_{st}^2$ from its posterior inverse Gamma distribution, $IG(a_{st}^*,b_{st}^*)$, with $a_{st}^*= a_0+(1/2)\sum_{i:S_i=s,T_i=t}D_i$, $b_{st}^*= b_0+(1/2)\sum_{i:S_i=s,T_i=t}\bbI\left(T_i=t\right)\left\{Y_i-m_{st}(\bfX_{st,i})\right\}^2$. 
Here $(a_0,b_0)$ and $(a^*,b^*)$ are shape and rate parameters for the prior and full conditional posterior distribution of $\sigma_{st}^2$, respectively. The full conditional of $\sigma_{st}^2$ only depends on the stratum-treatment group size, potential outcomes in that group, and the updated mean functions.
\item[3.] 
Update trees $\{\calT_{Z,1},\ldots,\calT_{Z,J_Z}\}$ and node parameters $\{\calM_{Z,1},\ldots,\calM_{Z,J_Z}\}$ for the $S$-model, using $Z_i$ as responses. Full conditional distributions of trees and node parameters in the $Z$-submodel depend on the latent variable, $Z_i$, covariate vector $\bfX_{Z,i}$, and prior distributions of trees and node parameters. Update mean functions $m_Z(\bfX_{Z,i})$ for $i=1,\ldots,n$, using the updated $\{\calT_{Z,1},\ldots,\calT_{Z,J_Z}\}$ and $\{\calM_{Z,1},\ldots,\calM_{Z,J_Z}\}$. These mean function values are needed for the sampling of $Z_i$ during data augmentation.
\item[4.] 
Update trees $\{\calT_{W,1},\ldots,\calT_{W,J_Z}\}$ and node parameters $\{\calM_{W,1},\ldots,\calM_{W,J_W}\}$ for the $S$-model, using $W_i$ as responses with $S_i=10,11$. Similar to the previous step, full conditional distributions of trees and node parameters in the $W$-submodel depend on the latent variable, $W_i$, latent stratum memberships, $S_i$, covariate vector $\bfX_{W,i}$, and prior distributions of trees and node parameters. Update $m_W(\bfX_{W,i})$ for $i=1,\dots,n$, using the updated $\{\calT_{W,1},\dots,\calT_{W,J_W}\}$ and $\{\calM_{W,1},\dots,\calM_{W,J_W}\}$. These mean function values are needed for the sampling of $W_i$ during data augmentation.
\item[5.] Update the stratum membership, $S_i$, for each participant. By Assumption \ref{asp3}, latent stratum memberships of certain subgroups of participants are directly ascertained, while the remaining participants require an update through data augmentation. Specifically,
\begin{itemize}
	\item[(a).] If $T_i=1$ and $D_i(1)=0$, then $S_i=00$;
	\item[(b).] If $T_i=0$ and $D_i(0)=1$, then $S_i=11$;
	\item[(c).] If $T_i=0$ and $D_i(0)=0$, then
	\begin{align*}
		p_{00,i}&=\bbP(S_i=00|\bullet)=\Phi\left(m_Z(\bfX_{Z,i})\right),\\
		p_{10,i}&=\bbP(S_i=10|\bullet)=\left\{1-\Phi\left(m_Z(\bfX_{Z,i})\right)\right\}\Phi\left(m_W(\bfX_{W,i})\right).
	\end{align*}
	Generate $\epsilon_i\sim \text{Bern}\left(p_{00,i}/(p_{00,i}+p_{10,i})\right)$. If $\epsilon_i=1$, set $S_i=00$; if $\epsilon_i=0$, set $S_i=10$. Here, full conditional probabilities, $p_{00,i}$ and $p_{10,i}$, do not involve potential outcomes, because the potential outcomes are not well-defined for participants assigned to the control arm ($T_i=0$) that died ($D_i(0)=0$). 
	\item[(d).] If $T_i=1$ and $D_i(1)=1$, then
	\begin{align*}
		p_{10,i}&=\bbP(S_i=10|\bullet)=\Phi\left(m_W(\bfX_{W,i})\right)\phi\left(Y_i;m_{101,i}(\bfX_{101,i}),\sigma_{101}^2\right),\\
		p_{11,i}&=\bbP(S_i=11|\bullet)=\left\{1-\Phi\left(m_W(\bfX_{W,i})\right)\right\}\phi\left(Y_i;m_{111,i}(\bfX_{111,i}),\sigma_{111}^2\right),
	\end{align*}
	where $\phi(y;m,\sigma^2)$ denotes the normal density with response $y$, mean $m$, and variance $\sigma^2$. Generate $\epsilon_i\sim \text{Bern}\left(p_{10,i}/(p_{10,i}+p_{11,i})\right)$. If $\epsilon_i=1$, set $S_i=10$; if $\epsilon_i=0$, set $S_i=11$. Here, potential outcomes are observable for participants assigned to the treatment arm ($T_i=1$) that survived ($D_i(1)=1$), such that full conditional probabilities, $p_{10,i}$ and $p_{11,i}$, depend on $Y_i$.
\end{itemize}
\item[6.] Update latent variable $Z_i$ for each participant by sampling from their truncated normal full conditional distributions:
\begin{align*}
	\{Z_i|m_Z(\bullet),\bfX_{Z,i},S_i\} &\sim \calN\left(m_Z(\bfX_{Z,i}),1\right)\bbI(Z_i\geq0), ~\text{if}~S_i=00,\\
	\{Z_i|m_Z(\bullet),\bfX_{Z,i},S_i\} &\sim \calN\left(m_Z(\bfX_{Z,i}),1\right)\bbI(Z_i<0), ~\text{if}~S_i=10~\text{or}~11.
\end{align*}
This is a data augmentation step that identifies participants in the never-survivor stratum in a specific iteration.
\item[7.] Update latent variable $W_i$ for each participant in stratum 10 and 11 by sampling from their truncated normal full conditional distributions:
\begin{align*}
	\{W_i|m_W(\bullet),\bfX_{W,i},S_i\} &\sim \calN\left(m_W(\bfX_{W,i}),1\right)\bbI(W_i\geq0), ~\text{if}~S_i=10,\\
	\{W_i|m_W(\bullet),\bfX_{W,i},S_i\} &\sim \calN\left(m_W(\bfX_{W,i}),1\right)\bbI(W_i<0), ~\text{if}~S_i=11.
\end{align*}
This is an additional data augmentation step that further identifies participants in the always-survivor stratum in a specific iteration.
\end{enumerate}

We initialize the proposed Gibbs sampler by first assigning participants to the three strata, where participants with directly identifiable stratum memberships are assigned directly, and those with stratum memberships that are not directly identifiable are randomly assigned to one of the possible strata according to their received treatments and survival status. For mean functions in the $Y$-model, $m_{st}(\bullet)$, initial estimates can be obtained using the BART model, e.g., using the {\bf bart} function from {\bf R} package {\bf dbarts}, given that all stratum memberships are fixed and the initial estimate for $\sigma_{st}^2$ can be simultaneously obtained from the fitting this initial BART model. For mean functions in the $S$-model, $m_Z(\bullet)$ and $m_W(\bullet)$, we fit parametric logistic regression models using indicators converted from initial stratum membership assignments and associated covariates; the resulting linear components are used as initial values for $m_Z(\bfX_{Z,i})$ and $m_W(\bfX_{W,i})$. Initial values for $Z_i$'s and $W_i$'s are generated from truncated normal distributions conditional on initial stratum membership assignments as well as initial estimates of $m_Z(\bfX_{Z,i})$ and $m_W(\bfX_{W,i})$ for all $i$. Hyperparameters of the inverse Gamma prior for $\sigma_{st}^2$ are chosen as $a_0=b_0=0.001$. The SACE and CSACE (for fixed value $\bfX=\bfx$) can then be calculated at each iteration of the Gibbs sampler, and the respective posterior distributions can be obtained after the sampling procedure is terminated. \textbf{R} code for implementing this Bayesian approach can be found at \href{https://github.com/erxc/BART-SACE-HTE}{https://github.com/erxc/BART-SACE-HTE}.

To compare the proposed Bayesian machine learning approach with its parametric counterparts in estimating the SACE and CSACE, we conducted simulations under two data generating processes and with two levels of total sample sizes. We find that (i) the proposed approach, where all mean functions in the $Y$- and $S$-model were specified nonparametrically using BART, outperformed parametric approaches in the estimation of both the SACE and CSACE, as measured by relative bias, root mean squared error and the overall precision in the estimation of heterogeneous effects \citep{Hill2011,Hu2021}; and that (ii) for estimating CSACE, the $Y$-model appears to play a more important role because using BART only in $Y$-model dominates using BART only in $S$-model in terms of both bias and efficiency. The details of the simulation study are presented in Web Appendix A2.

\section{Application to the ARDSNetwork ARMA study} \label{sec:app}

\subsection{Data}
The ARMA trial involved 861 participants with acute lung injury and acute respiratory distress syndrome who were randomized to receive mechanical ventilation with a volume of 12 mL per kilogram of predicted body weight ($T_i=0$) or a lower tidal volume ventilator strategy of 6 mL per kilogram of predicted body weight ($T_i=1$). We focus our analysis on the patient-centered non-mortality outcome variable DTRH with $180$ days as the maximum for those who survived; correspondingly, the principal strata based on potential survival status are defined at $180$ days. The DTRH captures important information to payer and health system stakeholders as a measure of health care utilization and to patients and their caregivers as it is associated with patients' long-term prognosis and health-related quality of life. We use $Y_i$ to denote the non-mortality outcome of participant $i$. In-hospital death events occurred in a substantial proportion (34.3\%) of enrolled trial participants, and more deaths were observed in the usual care group ($173/429=40.3\%$) than those in the treatment group ($146/473=30.9\%$) resulting in an absolute risk difference of $-9.4\%$. The study was motivated based on concerns that mechanical ventilation treatment using traditional tidal volumes of 10 to 15 mL per kilogram of body weight may cause stretch-induced lung injury in those with acute lung injury and acute respiratory distress syndrome \citep{brower2000acute}. We assume monotonicity such that the lower tidal volume does not lead to worse survival, and hence excludes the harmed stratum. We also exclude four participants who had one or more missing covariates. Summary statistics (means) of the non-mortality DTRH outcome and baseline covariates for the total sample of 857 enrolled participants by treatment arm $T_i$ and survival status $D_i$ are presented in Table \ref{tab:sum-stat}.

\begin{table}[htbp] 
\caption{Summary statistics of the key variables (means for numerical variables and proportions for binary indicators) in the ARMA trial.}\label{tab:sum-stat}
\par
\begin{center}
		\begin{tabular}{l c cccc} 
			\toprule 
			& All & $T_i=1,D_i=1$ & $T_i=1,D_i=0$ & $T_i=0,D_i=1$ & $T_i=0,D_i=0$ \\
			\cmidrule(lr){2-6}
			Sample size & $857$ & $303$ & $127$ & $260$ & $167$ \\
			\midrule
			DTRH ($Y_i$) & $-$ & 44.80 & $-$ & 47.94 & $-$ \\ [3pt]
			Age & 51.45 & 49.80 & 51.35 & 51.70 & 55.08 \\ 
			Sex (female) (\%) & 0.41 & 0.36 & 0.43 & 0.43 & 0.44 \\ 
			Race/ethnicity(\%) & & & & & \\ 
			~~White & 0.73 & 0.75 & 0.70 & 0.70 & 0.77 \\ 
			~~Non-White & 0.27 & 0.25 & 0.30 & 0.30 & 0.23 \\[3pt] 
			Tidal volume (ml) & 679.95 & 683.91 & 684.05 & 679.07 & 673.81 \\ 
			PEEP (cm water) & 8.40 & 8.33 & 9.05 & 8.11 & 8.50 \\ 
			PaO$_2$ (mm Hg) & 84.83 & 85.21 & 81.39 & 84.54 & 87.19 \\
			FiO$_2$ (mm Hg) & 0.63 & 0.61 & 0.66 & 0.61 & 0.68 \\ 
			PaCO$_2$ (mm Hg) & 36.32 & 36.43 & 36.08 & 37.18 & 34.99 \\ 
			PaO$_2$/FiO$_2$ (PtoF) & 149.00 & 155.46 & 135.50 & 153.24 & 140.98 \\ 
			AaDO$_2$ & 325.17 & 307.06 & 352.28 & 309.31 & 362.13 \\ 
			Arterial pH & 7.40 & 7.40 & 7.39 & 7.41 & 7.39 \\ [3pt]
			APACHE III & 82.52 & 76.04 & 92.87 & 76.95 & 95.11 \\ 
			Systolic BP & 97.88 & 100.55 & 94.89 & 99.80 & 92.34 \\ 
			Glasgow coma scale & 11.10 & 11.37 & 10.74 & 11.14 & 10.80 \\ 
			Platelet (count/nl) & 109.94 & 118.05 & 89.31 & 115.57 & 102.16 \\ 
			Creatine (mg/dl) & 1.18 & 1.10 & 1.15 & 1.14 & 1.40 \\ 
			Bilirubin (mg/dl) & 0.88 & 0.99 & 0.70 & 0.77 & 0.98 \\ 
			Vasopressors (\%) & 0.65 & 0.74 & 0.57 & 0.71 & 0.47 \\
			\bottomrule
		\end{tabular}
\end{center}
\end{table}

Based on discussions with clinical colleagues, we pre-selected 18 covariates that were measured at baseline, which can be broadly divided into three groups: (i) demographic information including age, sex, and race/ethnicity, (ii) respiratory measures including tidal volume in milliliter, positive end-expiratory pressure (PEEP) in centimeter water, fraction of inspired oxygen (FiO$_2$) in millimeter Hg, partial pressure of arterial carbon dioxide (PaCO$_2$) in millimeter Hg, partial pressure of arterial oxygen (PaO$_2$) in millimeter Hg,  the ratio of PaO$_2$ to FiO$_2$ (PtoF), the first alveolar-arterial oxygen gradient (AaDO$_2$), arterial pH, and (iii) physiological measures including the score of Acute Physiology, Age, and Chronic Health Evaluation (APACHE III), in addition to, the Glasgow coma scale score (Glasgow) as a measure of central nervous system failure, platelet count per nanoliter as a measure of coagulation, serum creatine in milligram per deciliter as a measure of renal function, bilirubin in milligram per deciliter as a measure of hepatic function, the use of vasopressors (indicating the need for blood pressure support), and systolic blood pressure (systolic BP) in millimeter Hg. Due to randomization, the baseline characteristics are comparable across different treatment groups. A descriptive comparison of outcomes among survivors suggests that participants receiving low tidal volume treatment appear to have shorter DTRH (average 44.80 days under low tidal volume treatment vs 47.94 days under the higher tidal volume ventilator strategy). To analyze DTRH outcomes subject to death truncation, We apply our proposed Bayesian machine learning method to estimate both the SACE and CSACE. 

\subsection{Implementation}
With the set of baseline covariates, we considered model \eqref{eq:outcome} for the potential outcomes. We standardized all continuous covariates to have zero mean and unit variance to improve the numerical stability. We implemented the Gibbs sampling procedure described in Section \ref{sec:inf}, by specifying the following priors. For the BART priors, the distribution on the choice of the splitting value at each internal node, the distribution of splitting value used at each internal node, and the probability that a node at given node-depth splits remained the same as previously described. For the distribution of terminal node values, we considered a five-fold cross-validation based on the set of $w\in\{1,2,3,4\}$ and $J\in\{50, 75, 100, 200\}$ (recall that $w$ and $J$ control the variance of the prior for the node values and the total number of trees), and found that $w=4$ and $J=50$ are associated with the best predictive performance of the outcome, and thus were adopted to generate our analytical results. Cross-validation results are shown in Web Figure A1 in  Web Appendix A3. We set $a_0=b_0=0.001$ for the Gamma priors of the variances. We ran the Markov Chain Monte Carlo procedure for 10,000 iterations and used the first 5,000 as burn-in. We obtained point estimates along with corresponding 95\% credible intervals of the SACE and CSACE based on draws from their respective posterior samples. The SACE here captures the average DTRH reduced under the low tidal volume treatment compared to the traditional volume treatment for the trial participants who classify as always-survivors.

\subsection{Survivor average causal effect and its conditional counterparts}

\subsubsection{Who would likely be the always-survivors?}\label{sec:match}

To characterize the CSACE, we first capture a trial subpopulation that is mostly likely being always-survivors. Because the stratum membership is not fully observed for all participants, capturing the subset of likely always-survivors is a practical decision that makes it feasible to study effect moderation on the non-mortality outcome. Under monotonicity, we differentiate the observed trial sample as follows: 
\begin{enumerate}
\item $\bbS_1$: subset of participants who are observed to be always-survivors; this set of participants are precisely those assigned to the treatment of traditional tidal volume and survived until the end of study;
\item $\bbS_0$: subset of participants who are observed to be never-survivors; this set of participants are precisely those assigned to the treatment of low tidal volume but died;
\item $\bbS_{(p,1)}$: subset of participants who are not in $\bbS_1$ or $\bbS_0$ but have a posterior probability of at least $p$ ($0<p<1$) to be always-survivors; this set of participants will be among those assigned to the treatment of traditional tidal volume but died prior to the end of study, or those assigned to the treatment of low tidal volume but survived until the end of study.
\end{enumerate}
In the ARMA trial, 30.3\%($=260/857$) of participants assigned to the treatment of traditional tidal volume survived until the day 180, while the posterior mean of the marginal proportion of always-survivors is estimated be to 60.9\%. This motivated us to consider using the set $\bbS_1\cup\bbS_{(p,1)}$ with $p=0.8$ to approximate the set of always-survivors. This choice of $p$ is given such that the proportion of this set matches the posterior mean of the marginal proportion of always-survivors returned by our chain. We then primarily focused on interpreting CSACE for participants who have at least 80\% posterior probability to belong to the always-survivor stratum. We further compared the posterior mean of $\bar{\Delta}_{CSACE}=N_{11}^{-1}\sum_{i=1}^{N_{11}}\Delta_{CSACE}(\bfX_i)$, where $N_{11}=|\bbS_1\cup\bbS_{(0.8,1)}|$, with the posterior mean of SACE, and found that they were identical. In particular, the posterior mean of SACE is $-23.87$ days and $\bar{\Delta}_{CSACE}=-23.87$ days. This post-hoc check ensures that $\bbS_1\cup\bbS_{(0.8,1)}$ is a reasonable approximation to the latent, always-survivor stratum, and confirms that the low tidal volume treatment leads to, on average, 24 days (95\% credible interval, 16.7-30.9 days) in reductions on DTRH among the always-survivors. That is, low tidal volume treatment led to substantial benefits regarding DTRH over the higher tidal volume treatment among the always-survivors subpopulation who are at a generally lower risk of death. This finding echoes the overall average treatment effect reported in the original trial analysis \citep{brower2000acute}.

\begin{table}[htbp] 
\caption{Estimated mean of covariates in two subpopulations and the posterior mean of absolute standardized differences by comparing between these two subpopulations. The second column corresponds to the mean of covariates in the identified likely always-survivors, and the third column corresponds to the posterior mean of the average covariate value in the latent always-survivor stratum.}\label{tab:compare-as}
\par
\begin{center}
		\begin{tabular}{l ccc} 
			\toprule 
			Covariates & $\bbS_1\cup\bbS_{(0.8,1)}$ & (Latent) always-Survivors & Posterior mean of ASD \\
			\midrule
			Age & 50.34 & 50.34 & 0.0019 \\
			Sex (female) (\%) & 0.386 & 0.385 & 0.0019 \\ 
			Race (White) (\%) & 0.271 & 0.270 & 0.0016 \\ 
			Tidal volume (ml) & 678.74 & 678.75 & 0.0012 \\
			PEEP (cm water) & 8.21 & 8.21 & 0.0015 \\ 
			PaO$_2$ (mm Hg) & 85.35 & 85.32 & 0.0014 \\ 
			FiO$_2$ (mm Hg) & 0.606 & 0.606 & 0.0026 \\  
			PaCO$_2$ (mm Hg) & 36.57 & 36.58 & 0.0012 \\ 
			PaO$_2$/FiO$_2$ (PtoF) & 155.04 & 154.91 & 0.0023 \\
			AaDO$_2$ & 306.40 & 306.75 & 0.0027 \\  
			Arterial pH & 7.4096 & 7.4096 & 0.0016 \\ 
			APACHE III & 76.39 & 76.40 & 0.0017 \\ 
			Systolic BP & 100.15 & 100.13 & 0.0012 \\ 
			Glasgow coma scale & 11.28 & 11.28 & 0.0016 \\ 
			Platelet (count/nl) & 117.26 & 117.43 & 0.0031 \\
			Creatine (mg/dl) & 1.117 & 1.115 & 0.0013 \\  
			Bilirubin (mg/dl) & 0.868 & 0.866 & 0.0014 \\  
			Vasopressors (\%) & 0.724 & 0.724 & 0.0015 \\ 
			\bottomrule
		\end{tabular}
\end{center}
\end{table}

To further assess the adequacy of using the likely always-survivor subpopulation for approximating the latent always-survivor subpopulation, we compare in Table \ref{tab:compare-as} the estimated means of covariates from these two subpopulations; the covariate means for the likely always-survivors are computed directly from $\bbS_1\cup\bbS_{(0.8,1)}$, whereas the covariate means for the latent always-survivor subpopulation are obtained from the posterior samples. In each iteration of our chain, we further estimate the absolute standardized difference (ASD) for each covariate and present its posterior mean in Table \ref{tab:compare-as}. The two subpopulations have almost identical covariate means and the posterior mean of ASD is below $1\%$ for each covariate (much lower than the usual 10\% cutoff suggested in \citet{austin2015moving} for observational studies). This finding reassures the plausibility of subsequent analyses based on the set of likely always-survivors.  

Before moving on to studying effect heterogeneity among the always-survivors, we provide some additional intuitions on who are the always-survivors in the ARMA study. Firstly, we obtain the variable importance plots generated from the BART models fitted for augmented latent variables, $Z$ and $W$, in Web Figures A2 and A3 of Web Appendix A3. From these plots, we observe that Systolic BP, AaDO$_2$, APACHE III, PtoF, FiO$_2$, Vasopressors, and Platelet are seven key variables of higher importance than the remaining variables. Secondly, we present in Web Table A2 the posterior mean of each covariate by latent stratum, along with the posterior mean of maximum pairwise ASD used in observational studies \citep{mccaffrey2013tutorial,Li2019}. The seven key variables identified by the variable importance plots, along with Arterial pH, correspond to the largest posterior mean ASD values (hence explaining differences between principal strata). From these results, we find that always-survivors are generally associated with having better health profiles, in terms of much lower APACHE III score, AaDO$_2$, FiO$_2$, higher platelet count, and younger in age. The always-survivors also show the highest percentage of vasopressors use with the highest average level of systolic BP.

\subsubsection{Visualizing conditional survivor average causal effects}
\begin{figure}[htbp]
	\centering
	\includegraphics[width=0.75\textwidth]{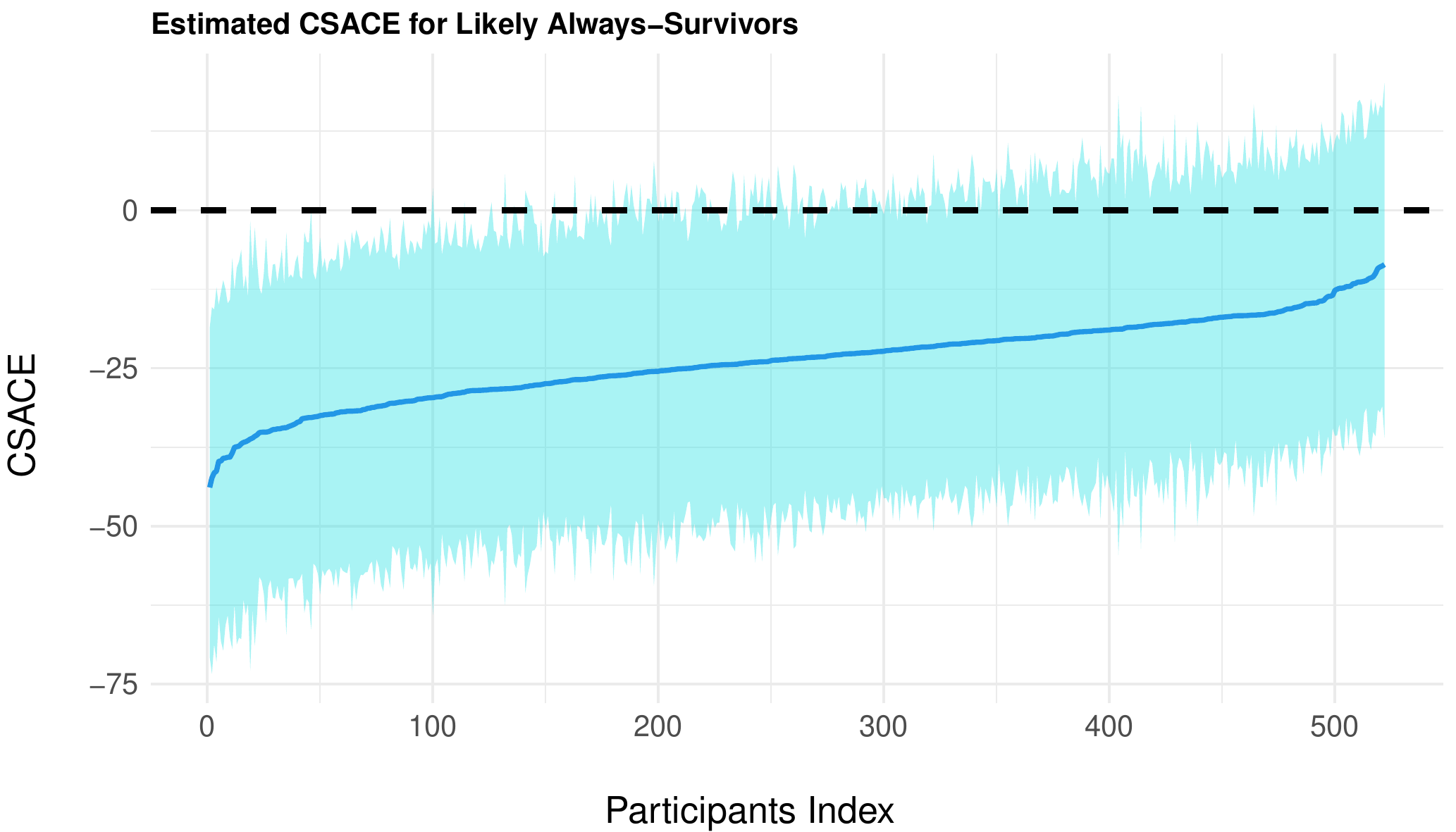}
	\caption{Posterior means of CSACE (darker blue) with corresponding 95\% credible intervals (lighter blue) for a total of 522 participants who are likely always-survivors (in $\bbS_1\cup\bbS_{(0.8,1)}$). A negative CSACE value indicates reduced DTRH under the low tidal volume treatment compared to the traditional tidal volume treatment, which is considered beneficial.}
	\label{fig:CSACE}
\end{figure}

Figure \ref{fig:CSACE} shows the posterior mean and 95\% credible intervals of $\Delta_{CSACE}(\bfX_i)$ for the 522 participants identified as likely always-survivors (in both $\bbS_1$ and $\bbS_{(0.8,1)}$). The plot indicates an overall benefit in terms of reducing the DTRH among those receiving low tidal volume treatment. But the conditional causal effects clearly differ to some degree, ranging from $-46.94$ to $-8.27$ days, suggesting heterogeneity in response to the low tidal volume treatment. A total of $37.8\%$ of identified likely always-survivors correspond to a credible interval excluding zero, which supports a strong, beneficial causal effect due to the low tidal volume treatment. 

\begin{figure}[htbp]
	\centering
	\includegraphics[width=0.8\textwidth]{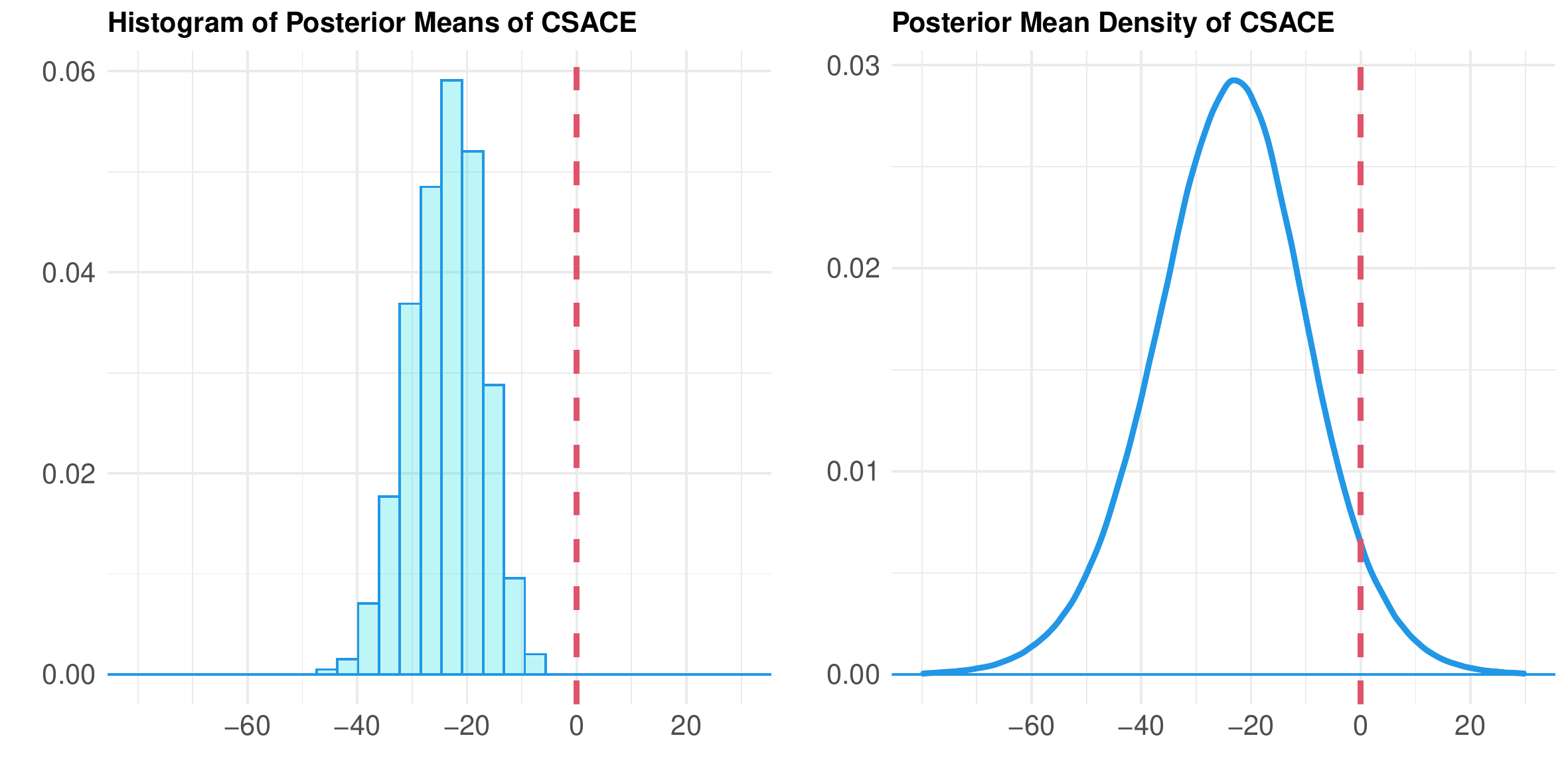}
	\caption{Left panel: histogram of posterior means of CSACE. The histogram is constructed using posterior means of CSACE of each likely always-survivor participants (in $\bbS_1\cup\bbS_{(0.8,1)}$). Right panel: posterior mean density of CSACE. The smooth estimate of the density function was computed as described in \eqref{eq:emp-den-csace}.}
	\label{fig:hist-csace}
\end{figure}

\citet{Henderson2018} provided several approaches for characterizing the degree of effect heterogeneity without truncation by death, and we apply their approaches to the ARMA trial for the identified likely always-survivors. To begin with, an alternative characterization of treatment effect heterogeneity can be achieved by examining the empirical distribution of the CSACEs over $\bbS_1\cup\bbS_{(0.8,1)}$, $H(u)=N_{11}^{-1}\sum_{i\in\bbS_1\cup\bbS_{(0.8,1)}}\bbI\left\{\Delta_{CSACE}(\bfX_i)\leq u\right\},$
which could be directly estimated by
\begin{align} \label{eq:emp-dist-csace}
	\hat{H}(u)=\frac{1}{N_{11}}\sum_{i\in\bbS_1\cup\bbS_{(0.8,1)}}\bbP\left\{\Delta_{CSACE}(\bfX_i)\leq u|\text{Obs. Data},\bftheta\right\}.
\end{align}
To better visualize the spread of CSACE over $\bbS_1\cup\bbS_{(0.8,1)}$, we estimate the density function associated with \eqref{eq:emp-dist-csace} by computing the posterior mean of a kernel function $K_\lambda(\bullet)$:
\begin{align} \label{eq:emp-den-csace}
	\hat{h}(u)=\frac{1}{N_{11}}\sum_{i\in\bbS_1\cup\bbS_{(0.8,1)}}\bbE\left\{K_\lambda\left(u-\Delta_{CSACE}(\bfX_i)\right)|\text{Obs. Data},\bftheta\right\}.
\end{align}
The bandwidth $\lambda$ is set as $\{0.9\times\min(\hat{\sigma}_{CSACE},\widehat{IQR}_{CSACE})\}/(1.34\times N_{11}^{1/5})$, where $\hat{\sigma}_{CSACE}$ and $\widehat{IQR}_{CSACE}$ are posteriors means of the standard deviation and inter-quartile range of $\Delta_{CSACE}(\bfX_i)$. The left panel of Figure \ref{fig:hist-csace} presents a histogram of the posterior means of CSACE for each participant in $\bbS_1\cup\bbS_{(0.8,1)}$, and the right panel is the estimated posterior mean empirical density (the average of $\hat{h}(u)$ obtained at each MCMC iteration), which refers to the estimate of the entire distribution of the underlying treatment effects among $\bbS_1\cup\bbS_{(0.8,1)}$. The variation in the treatment effect suggested by the right panel is larger than that by the left panel, matching the intuition that the variance of conditional means is often smaller than the individual variation. Nonetheless, the estimated CSACEs were primarily negative via either visualization technique, leading to converging evidence. Thus, we conclude that the low tidal volume treatment leads to shorter DTRH, and the greatest reduction according to the posterior means of CSACE reaches over 45 days. 

\subsubsection{Quantifying heterogeneity in conditional survivor average causal effects}

In addition to visualization, we apply two metrics considered in \citet{Henderson2018} to the ARMA trial for quantifying the degree of heterogeneity in the estimated CSACE among the likely always-survivors. First, the existence of heterogeneity can be quantified using the posterior probabilities of the differential survivor causal effect for each participant $i\in\bbS_1\cup\bbS_{(0.8,1)}$, defined as 
\begin{align*} 
	\calD_i = \bbP\left\{\Delta_{CSACE}(\bfX_i)\leq \bar{\Delta}_{CSACE}|\text{Obs. Data},\bftheta\right\},
\end{align*}
along with the absolute differential survivor causal effect,
\begin{align*} 
	\calD_i^* = \max\left\{1-2\calD_i,2\calD_i-1\right\},
\end{align*}
where $\bar{\Delta}_{CSACE}$ is the average of the CSACE among $\bbS_1\cup\bbS_{(0.8,1)}$. Notice that in the ARMA trial application, we have verified that $\bar{\Delta}_{CSACE}\approx\Delta_{SACE}$, and therefore the differential survivor causal effect can be approximately equivalently defined as $\calD_i = \bbP\left\{\Delta_{CSACE}(\bfX_i)\leq {\Delta}_{SACE}|\text{Obs. Data},\bftheta\right\}$.

The differential survivor causal effect, $\calD_i$, is a measure of the evidence that the CSACE, $\Delta_{CSACE}(\bfX_i)$, is less than or equal to the average of CSACE among the set of likely always-survivors, and thus, we should expect both high and low values of $\calD_i$ in settings where non-negligible heterogeneity of treatment effects exists. The closely-related quantity, $\calD_i^*$, approaches 1 as the value of $\calD_i$ approaches either 0 or 1, and $\calD_i^*=0$ when $\calD_i=1/2$. For a given participant $i\in\bbS_1\cup\bbS_{(0.8,1)}$, we therefore consider there to be strong evidence of heterogeneity in CSACEs if $\calD_i^*>0.9$ (equivalently, if $\calD_i<0.05$ or $\calD_i>0.95$), moderate evidence of heterogeneity provided that $\calD_i^*>0.8$ (equivalently, if $\calD_i<0.1$ or $\calD_i>0.9$), and mild evidence of heterogeneity if $\calD_i^*>0.7$ (equivalently, if $\calD_i<0.15$ or $\calD_i>0.85$). In the simulation study by \citet{Henderson2018} without truncation by death, for cases with treatment effect homogeneity, they found that the proportion of participants exhibiting high values of the $\calD_i^*$ should, ideally, be zero or quite close to zero. For this reason, the proportion of participants with $\calD_i^*>0.9$ can potentially be a useful summary measure for detecting heterogeneity in CSACEs. 
In the ARMA trial, approximately 0.4\% of participants had strong evidence of heterogeneity in CSACEs (i.e. $\calD_i^*>0.9$), approximately 1.3\% of participants had moderate evidence of heterogeneity (i.e. $\calD_i^*>0.8$), and approximately 6.1\% of participants had mild evidence of heterogeneity (i.e. $\calD_i^*>0.7$). Web Figure A4 in Web Appendix A3 presents the histogram and density describing the distribution of $\calD_i^*$.

\begin{table}[htbp] 
	\caption{Tabulation of proportions of participants in $\bbS_1\cup\bbS_{(0.8,1)}$ benefiting from the low tidal volume treatment to different degrees.}\label{tab:dsce-prop}
	\par
	\begin{center}
			\begin{tabular}{cc} 
				\toprule 
				Benefiting degree & Proportion (\%) among $\bbS_1\cup\bbS_{(0.8,1)}$\\
				\midrule
				$\bbP\{\Delta_{CSACE}(\bfX_i)<0|\text{Obs. Data},\bftheta\}>0.99$ & 19.0 \\
				$\bbP\{\Delta_{CSACE}(\bfX_i)<0|\text{Obs. Data},\bftheta\}>0.95$ & 68.4 \\
				$\bbP\{\Delta_{CSACE}(\bfX_i)<0|\text{Obs. Data},\bftheta\}>0.9$ & 88.9 \\
				$\bbP\{\Delta_{CSACE}(\bfX_i)<0|\text{Obs. Data},\bftheta\}>0.8$ & 98.5 \\
				\bottomrule
			\end{tabular}
	\end{center}
\end{table}

Second, the heterogeneity in CSACEs can also be assessed via the proportion of always-survivors benefiting from the treatment \citep{Henderson2018}, where we directly infer the number of participants benefiting from the low tidal volume treatment from the set of participants who are likely always-survivors. In specific, the proportion of always-survivors benefiting from the low tidal volume treatment can be defined as
\begin{align*}
	\calQ = \frac{1}{N_{11}}\sum_{i\in\bbS_1\cup\bbS_{(0.8,1)}}\bbI\left\{\Delta_{CSACE}(\bfX_i)<0\right\}.
\end{align*}
The posterior mean of $\calQ$ is an average of the posterior probabilities of treatment benefit, $\hat{q}_i=\bbP\{\Delta_{CSACE}(\bfX_i)<0|\text{Obs. Data},\bftheta\}$, which summarizes the treatment benefit of a participant from a probabilistic perspective. Trial participants who are more likely to benefit from the low tidal volume treatment will have higher chances of a negative CSACE. 
A tabulation of participants among the likely always-survivors according to their likelihood of benefiting from the low tidal volume treatment is presented in Table \ref{tab:dsce-prop}, where 68.4\% of participants in $\bbS_1\cup\bbS_{(0.8,1)}$ exhibit a posterior probability of benefiting from the low tidal volume treatment greater than 0.95, and 88.9\% exhibit a posterior probability of benefiting from the low tidal volume treatment greater than 0.9. Web Figure A5 in Web Appendix A3 presents the histogram and density describing the distribution of $\hat{q}_i$.


\subsection{Exploring effect moderation}\label{sec:subgroup}

\begin{figure}[htbp]
	\centering
	\includegraphics[width=0.75\textwidth]{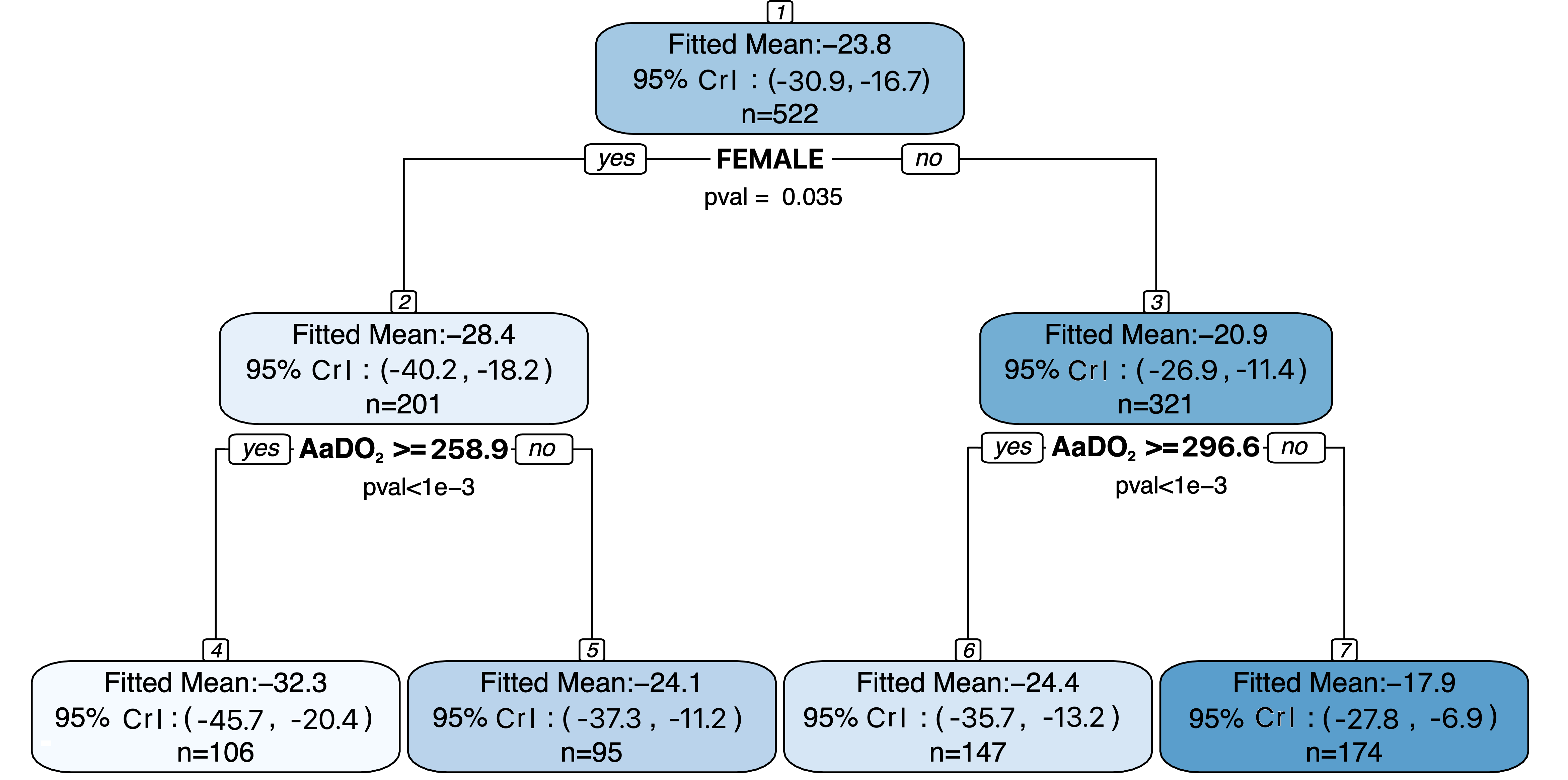}
	\caption{Final CART model fit to the posterior mean DTRHs (in days) between the low tidal volume treatment and the traditional tidal volume treatment. Values in each node correspond to the posterior mean and 95\% credible intervals for the average CSACE for the subgroup of individuals represented in that node.}
	\label{fig:tree-plot}
\end{figure}

We adopt the Bayesian ``fit-the-fit'' strategy \citep{Hahn2020} to explore the relationship between CSACEs and covariates among the likely always-survivors. This approach amounts to first applying our proposed method to estimate the CSACEs for each likely always-survivor (in $\bbS_1\cup\bbS_{(0.8,1)}$) and then, using these estimated CSACEs as a new response variable in an exploratory analysis to identify important effect moderators and possible subgroups defined by such effect moderators. Specifically, in our exploratory analysis, a classification and regression tree (CART) model was used to regress the posterior means of the CSACE on the covariates. We fit a sequence of CART models, with covariates (standardized to have zero mean and unit variance) sequentially added to the CART model in a stepwise manner to improve the model fit measured by $R^2$. At each step, the variable leading to the largest $R^2$ improvement was selected into the model, and, the procedure was terminated when the percent improvement in $R^2$ was less than 1\%. Results showed that covariates with the five largest estimated standardized coefficients in absolute value were (from high to low): AaDO$_2$, sex, FIO$_2$, PtoF, and systolic BP. Subgroup treatment effects were estimated by averaging CSACEs among individuals falling into each node of the final CART model, and the branch decision rules suggest final combination rules of covariates. Figure \ref{fig:tree-plot} presents the final tree estimates based on the top two covariates that are the main drivers of the heterogeneity in CSACE, where the final $R^2$ between the tree fit, and the posterior mean CSACE is 78.9\%. The 95\% credible interval (CrI) of each subgroup causal effect is obtained by projecting the posterior draws of CSACE onto the predictive space of the final CART fit, and hence comes with a valid Bayesian uncertainty interpretation \citep{Woody2021}.

\begin{figure}[htbp]
	\centering
	\includegraphics[width=0.85\textwidth]{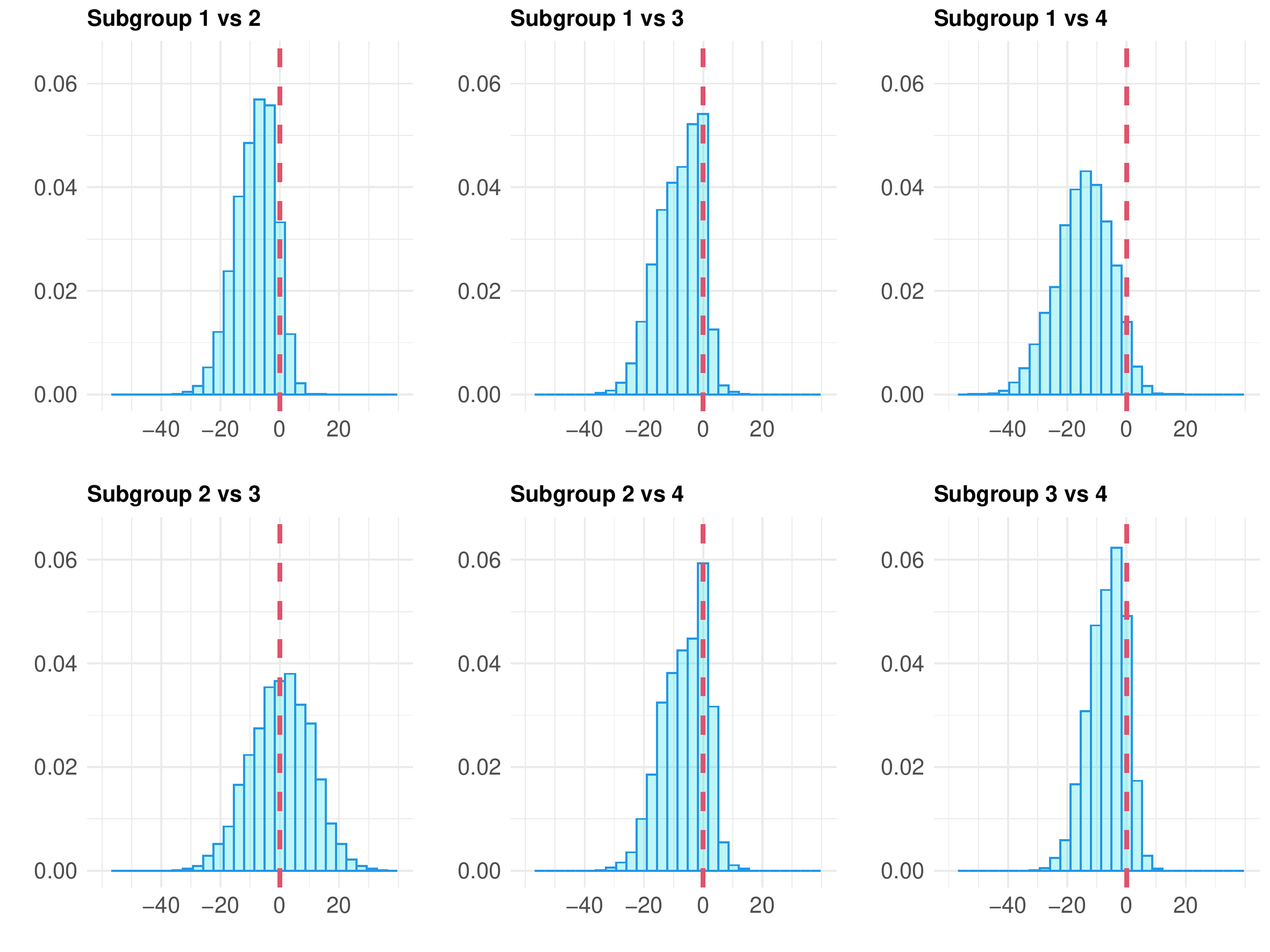}
	\caption{Posterior distributions of the difference in treatment effects between any two always-survivor subgroups. Subgroup 1: female always-survivors with $\text{AaDO}_2 \geq 258.9$; Subgroup 2: female always-survivors with $\text{AaDO}_2 < 258.9$; Subgroup 3: male always-survivors with $\text{AaDO}_2 \geq 296.6$; and Subgroup 4: male always-survivors with $\text{AaDO}_2 < 296.6$.}
	\label{fig:hist-diff}
\end{figure}

In Figure \ref{fig:tree-plot}, the first splitting variable was sex. Female always-survivors had approximately 28.4 (95\% CrI: 18.2–40.2) days shorter in DTRH on average under the low tidal volume treatment, whereas male always-survivors had approximately 20.9 (95\% CrI: 11.4–26.9) days shorter in DTRH on average under the low tidal volume treatment. The second level of variable splitting by the value of AaDO$_2$, the first alveolar-arterial oxygen gradient, provided further resolution on the magnitude of the treatment benefit for participants. The most beneficial subgroup was female always-survivors with $\text{AaDO}_2 \geq 258.9$, where the average reduction in DTRH is 32.3 (95\% CrI: 20.4–45.7) under the low tidal volume treatment. Among male always-survivors, those with $\text{AaDO}_2 < 296.6$ experience treatment benefit from the low tidal volume treatment with an average DTRH of approximately 17.9 (95\% CrI: 6.9–27.8) days shorter; in comparison, male always-survivors with $\text{AaDO}_2 \geq 296.6$ experience even greater treatment benefit from the low tidal volume treatment with an average DTRH of approximately 24.4 (95\% CrI: 13.2–35.7). Concordant with Figure \ref{fig:tree-plot}, we also visualize the posterior distribution of each pairwise difference in the subgroup causal effects in Figure \ref{fig:hist-diff}. It is apparent that the female always-survivors with $\text{AaDO}_2 \geq 258.9$ have the largest benefit from the low tidal volume treatment, as the majority of posterior mass in treatment effect difference is below zero when contrasting this subgroup to the others.

\begin{figure}[htbp]
\centering
\includegraphics[width=0.85\textwidth]{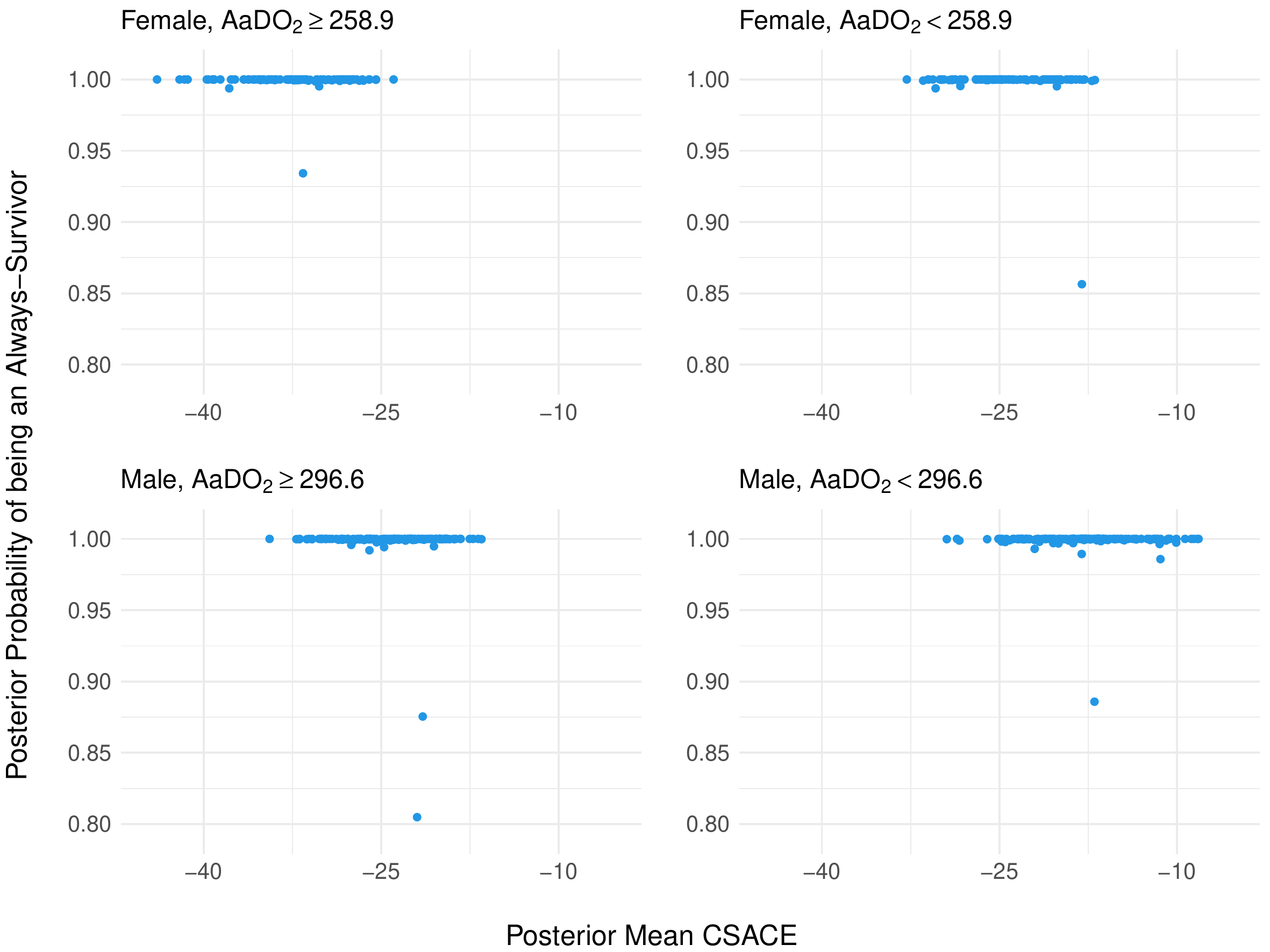}
\caption{Scatter plots of the posterior probability of being in the always-survivor stratum against the posterior mean CSACE by each subgroup.}
\label{fig:subgroup-scat}
\end{figure}

Finally, we also explore the patterns between the estimated CSACE and the posterior probability of being an always-survivor within each covariate-defined subgroup in Figure \ref{fig:subgroup-scat}. This exploration reveals a slight tendency that a subgroup with larger treatment benefits may have higher probability of being always-survivors. For example, among the set of likely always-survivors (Section \ref{sec:match}), there are only four participants with estimated posterior probability of being an always-survivor lower than 90\%---one female with $\text{AaDO}_2 < 258.9$, two males with $\text{AaDO}_2 \geq 296.6$, and one male with $\text{AaDO}_2 < 296.6$. The most beneficial subgroup correspond to the least uncertainty in being always-survivors; that is, except for one participant, all female with $\text{AaDO}_2 \geq 258.9$ have estimated posterior probability of being an always-survivor being at least $99\%$.



Overall, our exploratory analyses indicate that the reduction in DTRH was greatest among female always-survivors with $\text{AaDO}_2 \geq 258.9$ at baseline, and is smallest among male always-survivors with $\text{AaDO}_2 < 296.6$. The effect among females is consistent with prior findings in existing observational studies. For instance, a study from the Large Observational Study to Understand the Global Impact of Severe Acute Respiratory Failure (LUNG SAFE) \citep{McNicholas2019}, which is an international, multicenter, prospective cohort study, conducted for four consecutive weeks in the winter of 2014 in a convenience sample of 459 ICUs from 50 countries across six continents, and found that surviving females had a shorter duration of invasive mechanical ventilation and reduced length of stay compared with males. Second, participants with more severe acute respiratory disease syndrome have lower PaO$_2$:FiO$_2$ ratios and larger $\text{AaDO}_2$ gradients \citep{Helmholz1979}. Thus, there is some speculation that individuals with severe acute respiratory distress syndrome may be more likely to benefit from the intervention, whereas those with smaller gradients would be more strongly associated with poor clinical outcomes, such as death, or in our context, discharge to a long term acute care hospital, skilled nursing facility, or hospice, thereby delaying time to get home. In other words, the always-survivors with relatively higher $\text{AaDO}_2$ had more ``opportunity to benefit'' \citep{goligher2021effect}. 
Taken together, female always-survivors appear to benefit more from the low tidal volume treatment than their male counterparts. Thus, while the exact mechanisms may not be clear, our findings do seem plausible and directly engage with current debates in the treatment of acute lung injury and acute respiratory disease syndrome and the associated research literature \citep{Del-Sorbo2017,Fan2017,Shen2019}.

\section{Discussion} \label{sec:con}


Recent advancements in Bayesian machine learning have provided important tools to flexibly specify the outcome model to reduce the potential estimation bias that occurs when estimating the average treatment effect, and has enabled researchers to estimate heterogeneous causal effects among the study population. This article advances the application of BART to quantify the SACE and CSACE within the principal stratification framework when a non-mortality outcome is subject to truncation by death, and thus opens the door to a wide range of causal discoveries that could inform individualized care delivery in the motivating critical care use case. We applied our proposed approach to operationalize considerations for \emph{exploratory} heterogeneity of treatment effect analysis among the likely always-survivors in the ARMA trial, and identified key effect moderators using a data-driven approach that aligns with several clinical prior findings, as we explicate in Section \ref{sec:subgroup}. 

Beyond effect moderation due to sex and AaDO$_2$ we found in our analysis of the ARMA trial, the Bayesian ``fit-the-fit'' strategy that we employed also identified pressure of arterial oxygen, the ratio of PaO$_2$ to FiO$_2$, and systolic blood pressure as three additional factors that weakly moderate the causal effects among the always-survivors. However, the subgroup structure with more effect moderators necessarily becomes more complex and less interpretable, however, it is worth noting that these findings all still align with the clinical literature. We, therefore, decided to prioritize the top two effect moderators in our final exploratory analysis but fully acknowledge the value of future work for better synthesizing more than two effect moderators to generate interpretable subgroup findings. To the best of our knowledge, this is the first study that employed Bayesian machine learning tools to study effect moderation for mechanical ventilation treatments among the always-survivors population in a critical care intervention study. The investigation of the true causal mechanisms of such effect moderation will be left for future studies and necessitates structured engagement with a wider set of clinical colleagues.

In exploring the variation among the CSACE estimates, we have implemented a decision to first identify the likely always-survivors, which include $260$ survivors receiving the high tidal volume treatment and a subset of survivors ($262$ out of $303$) receiving the low tidal volume treatment but having the highest posterior probability of being an always-survivor. In Section \ref{sec:match}, we have compared the baseline characteristics among the likely always-survivors and those among the latent always-survivors (generated through our posterior sampling algorithm), and found no systematic difference. The SACE estimates are also identical between these two sets of participants, suggesting no clear evidence against the adequacy of using the likely always-survivors to approximate the latent always-survivors. Subsequently, pursuing the CSACE analysis with the likely always-survivors is based on two practical considerations. First, having a tangible subset of participants helps us directly study the variation in treatment effect for the non-mortality outcome (response heterogeneity), without the complications due to variation in the conditional probability of being an always-survivor (membership heterogeneity). Had we estimated CSACE based on covariates of the entire trial, it would be necessary to disentangle response heterogeneity from membership heterogeneity, which is challenging. Second, we recognize that an alternative approach is to focus on the $260$ survivors receiving the high tidal volume treatment. Under Assumptions \ref{asp2} and \ref{asp3}, this smaller subset is always a valid approximation to the latent always-survivor subpopulation. However, this alternative approach comes at a cost of substantially reduced sample size for exploring heterogeneity of treatment effect. Since we did not find systematic differences between the likely always-survivors and the latent always-survivors, we considered an analysis with the largest possible sample size. In cases where the stratum membership can not be easily predicted and hence the use of likely always-survivors may be questionable, it would then be preferable to focus the analysis on the smaller set of survivors under the usual care condition.

To estimate the SACE for non-mortality outcomes truncated by death, typically, both structural assumptions and parametric modeling assumptions are invoked. The structural assumptions are necessary to identify the causal parameter with observed data, whereas the parametric assumptions are useful in modeling the observed data and summarizing information from observed data. Under the principal stratification framework, the proposed Bayesian machine learning approach differs from the existing methods in that we considered a finite mixture of BART $Y$-models (with mixture probability also given by nested Probit BART models), rather than a finite mixture of fully parametric $Y$-models, thus relaxing some of the parametric modeling assumptions. In simpler settings without any intermediate outcomes, the BART approach has shown to be a flexible and robust tool to estimate the average treatment effect and its conditional counterpart with minimum bias and high precision \citep{Hill2011,dorie2019automated,Hahn2020,Hu2021}. Under this perspective, our work represents a generalization of the BART approach to additionally account for an intermediate variable through a mixture model framework. 
While relaxing the parametric assumptions, our approach still maintains standard structural assumptions to estimate the SACE. The SUTVA and randomization assumptions are generally plausible in applications to randomized trials, but the monotonicity assumption may not always be plausible, such as in non-inferiority or comparative effectiveness trials where there is an active comparator. In this case, one potential solution is to allow for an additional harmed strata \citep{Zhang2009jasa} by extending the nested Probit BART with another layer and including an additional $Y$-model for the harmed population under the usual care condition. Alternatively, it may be interesting to consider the monotone probit BART for mixture weights similar to \citet{Papakostas2021} to reflect a stochastic monotonicity constraint that one active treatment does not mitigate the risk of mortality compared to the other. While theoretically appealing, such extended approach may be over-parameterized and lead to semiparametric mixture models that are only weakly identified in the sense that the posterior distributions of SACE and CSACE remain flat around the region of highest density. Of note, integrating BART into the mixture model framework for principal stratification analysis is not the only approach to address heterogeneity of treatment effect under truncation by death. In future work, it would be interesting to compare the proposed BART approach with alternative Bayesian nonparametric priors, such as the dependent Dirichlet process-Gaussian process prior \citep{xu2016bayesian,roy2017bayesian,Xu2022}, for estimating CSACE. In addition, the extent to which alternative identification strategies \citep{Hayden2005,ding2017principal} might improve the current mixture model framework to estimate CSACE can also form the scope of further research.

A relevant extension of our approach is to view both DTRH and time-to-death as two time-to-event outcomes under the semi-competing risks framework. Under this framework, a continuous-time principal stratification approach has been developed in \citet{comment2019survivor}, \citet{Xu2022} and \citet{Nevo2021} to define always-survivors up to each follow-up time point, based on which a time-varying version of SACE (TV-SACE) is proposed. Our SACE estimand defined in Section \ref{sec:not-asp} can thus be viewed as a ``snapshot'' version of TV-SACE at $u=180$ days; and we focus on a snapshot version of principal stratification to facilitate the exploration of CSACE without addressing complications due to temporal heterogeneity of treatment effect. As a supplementary analysis, we implemented the Bayesian nonparametric approach of \citet{Xu2022} using the \textbf{BaySemiCompeting} R package to estimate the TV-SACE from the ARMA data; the details are summarized in Web Appendix A5. The results indicate that low tidal volume treatment strategy leads to consistently higher chances of returning home prior to day $u$ among the always-survivors up to day $u$, for each $0<u\leq 180$. However, a full development of formal methodology for estimation and interpretation of time-varying conditional survivor average causal effect (TV-CSACE), possibly through BART for time-to-event outcomes \citep{Henderson2018,Hu2021}, can be a fruitful direction for further investigation.

\begin{acks}[Acknowledgments]
The authors would like to extend their gratitude, without any implication for any errors in reporting or interpretation, to Drs. Douglas Hayden, B. Taylor Thompson, Scott Halpern, and Nadir Yehya for assistance with various questions during the development of this manuscript. 
\end{acks}

\begin{funding}
Research in this article was partially supported by the Patient-Centered Outcomes Research Institute\textsuperscript{\textregistered} (PCORI\textsuperscript{\textregistered} Awards ME-2020C1-19220 to M.O.H. and ME-2020C3-21072 to F.L). M.O.H. is funded by the United States National Institutes of Health (NIH), National Heart, Lung, and Blood Institute (NHLBI, grant number R00-HL141678). X.C., F.L., and M.O.H. are funded by the NIH/NHLBI (grant number R01-HL168202). All statements in this report, including its findings and conclusions, are solely those of the authors and do not necessarily represent the views of the NIH or PCORI\textsuperscript{\textregistered} or its Board of Governors or Methodology Committee.
\end{funding}

\begin{supplement}
\stitle{Web Appendix of ``A Bayesian Machine Learning Approach for Estimating Heterogeneous Survivor Causal Effects: Applications to a Critical Care Trial''}
\sdescription{The Web Appendix contains the Gibbs sampler of the parametric model (A1), a Monte Carlo simulation study (A2), related Web Figures (A3), and two aforementioned sensitivity analyses (A4 and A5).}
\end{supplement}


\bibliographystyle{imsart-nameyear} 
\bibliography{BART_SACE_HTE}       

\begin{thebibliography}{54}

\bibitem[\protect\citeauthoryear{Albert and Chib}{1993}]{albert1993bayesian}
\begin{barticle}[author]
\bauthor{\bsnm{Albert},~\bfnm{James~H}\binits{J.~H.}} \AND
  \bauthor{\bsnm{Chib},~\bfnm{Siddhartha}\binits{S.}}
(\byear{1993}).
\btitle{Bayesian analysis of binary and polychotomous response data}.
\bjournal{Journal of the American Statistical Association}
\bvolume{88}
\bpages{669-679}.
\end{barticle}
\endbibitem

\bibitem[\protect\citeauthoryear{Austin and Stuart}{2015}]{austin2015moving}
\begin{barticle}[author]
\bauthor{\bsnm{Austin},~\bfnm{Peter~C}\binits{P.~C.}} \AND
  \bauthor{\bsnm{Stuart},~\bfnm{Elizabeth~A}\binits{E.~A.}}
(\byear{2015}).
\btitle{Moving towards best practice when using inverse probability of
  treatment weighting (IPTW) using the propensity score to estimate causal
  treatment effects in observational studies}.
\bjournal{Statistics in Medicine}
\bvolume{34}
\bpages{3661-3679}.
\end{barticle}
\endbibitem

\bibitem[\protect\citeauthoryear{Bargagli-Stoffi, De~Witte and
  Gnecco}{2022}]{Bargagli2022}
\begin{barticle}[author]
\bauthor{\bsnm{Bargagli-Stoffi},~\bfnm{Falco~J.}\binits{F.~J.}},
  \bauthor{\bsnm{De~Witte},~\bfnm{Kristof}\binits{K.}} \AND
  \bauthor{\bsnm{Gnecco},~\bfnm{Giorgio}\binits{G.}}
(\byear{2022}).
\btitle{Heterogeneous causal effects with imperfect compliance: A Bayesian
  machine learning approach}.
\bjournal{The Annals of Applied Statistics}
\bvolume{0}
\bpages{1-19}.
\end{barticle}
\endbibitem

\bibitem[\protect\citeauthoryear{Bia, Mattei and Mercatanti}{2021}]{Bia2021}
\begin{barticle}[author]
\bauthor{\bsnm{Bia},~\bfnm{Michela}\binits{M.}},
  \bauthor{\bsnm{Mattei},~\bfnm{Alessandra}\binits{A.}} \AND
  \bauthor{\bsnm{Mercatanti},~\bfnm{Andrea}\binits{A.}}
(\byear{2021}).
\btitle{Assessing causal effects in a longitudinal observational study with
  ``truncated'' outcomes due to unemployment and nonignorable missing data}.
\bjournal{Journal of Business \& Economic Statistics}
\bvolume{0}
\bpages{1-12}.
\bdoi{10.1080/07350015.2020.1862672}
\end{barticle}
\endbibitem

\bibitem[\protect\citeauthoryear{Brower et~al.}{2000}]{brower2000acute}
\begin{barticle}[author]
\bauthor{\bsnm{Brower},~\bfnm{Roy~G}\binits{R.~G.}},
  \bauthor{\bsnm{Matthay},~\bfnm{Michael~A}\binits{M.~A.}},
  \bauthor{\bsnm{Morris},~\bfnm{Alan}\binits{A.}},
  \bauthor{\bsnm{Schoenfeld},~\bfnm{David}\binits{D.}},
  \bauthor{\bsnm{Thompson},~\bfnm{B~Taylor}\binits{B.~T.}},
  \bauthor{\bsnm{Wheeler},~\bfnm{Arthur}\binits{A.}} \betal{et~al.}
(\byear{2000}).
\btitle{Acute Respiratory Distress Syndrome Network. Ventilation with lower
  tidal volumes as compared with traditional tidal volumes for acute lung
  injury and the acute respiratory distress syndrome}.
\bjournal{New England Journal of Medicine}
\bvolume{342}
\bpages{1301-1308}.
\end{barticle}
\endbibitem

\bibitem[\protect\citeauthoryear{Chiba and VanderWeele}{2011}]{Chiba2011}
\begin{barticle}[author]
\bauthor{\bsnm{Chiba},~\bfnm{Yasutaka}\binits{Y.}} \AND
  \bauthor{\bsnm{VanderWeele},~\bfnm{Tyler~J.}\binits{T.~J.}}
(\byear{2011}).
\btitle{{A simple method for principal strata effects when the outcome has been
  truncated due to death}}.
\bjournal{American Journal of Epidemiology}
\bvolume{173}
\bpages{745-751}.
\bdoi{10.1093/aje/kwq418}
\end{barticle}
\endbibitem

\bibitem[\protect\citeauthoryear{Chipman, George and
  McCulloch}{2010}]{Chipman2010}
\begin{barticle}[author]
\bauthor{\bsnm{Chipman},~\bfnm{Hugh~A.}\binits{H.~A.}},
  \bauthor{\bsnm{George},~\bfnm{Edward~I.}\binits{E.~I.}} \AND
  \bauthor{\bsnm{McCulloch},~\bfnm{Robert~E.}\binits{R.~E.}}
(\byear{2010}).
\btitle{{BART: Bayesian additive regression trees}}.
\bjournal{The Annals of Applied Statistics}
\bvolume{4}
\bpages{266-298}.
\bdoi{10.1214/09-AOAS285}
\end{barticle}
\endbibitem

\bibitem[\protect\citeauthoryear{Comment et~al.}{2019}]{comment2019survivor}
\begin{barticle}[author]
\bauthor{\bsnm{Comment},~\bfnm{Leah}\binits{L.}},
  \bauthor{\bsnm{Mealli},~\bfnm{Fabrizia}\binits{F.}},
  \bauthor{\bsnm{Haneuse},~\bfnm{Sebastien}\binits{S.}} \AND
  \bauthor{\bsnm{Zigler},~\bfnm{Corwin}\binits{C.}}
(\byear{2019}).
\btitle{Survivor average causal effects for continuous time: a principal
  stratification approach to causal inference with semicompeting risks}.
\bjournal{arXiv preprint arXiv:1902.09304}
\bpages{1-28}.
\end{barticle}
\endbibitem

\bibitem[\protect\citeauthoryear{Del~Sorbo et~al.}{2017}]{Del-Sorbo2017}
\begin{barticle}[author]
\bauthor{\bsnm{Del~Sorbo},~\bfnm{Lorenzo}\binits{L.}},
  \bauthor{\bsnm{Goligher},~\bfnm{Ewan~C.}\binits{E.~C.}},
  \bauthor{\bsnm{McAuley},~\bfnm{Daniel~F.}\binits{D.~F.}},
  \bauthor{\bsnm{Rubenfeld},~\bfnm{Gordon~D.}\binits{G.~D.}},
  \bauthor{\bsnm{Brochard},~\bfnm{Laurent~J.}\binits{L.~J.}},
  \bauthor{\bsnm{Gattinoni},~\bfnm{Luciano}\binits{L.}},
  \bauthor{\bsnm{Slutsky},~\bfnm{Arthur~S.}\binits{A.~S.}} \AND
  \bauthor{\bsnm{Fan},~\bfnm{Eddy}\binits{E.}}
(\byear{2017}).
\btitle{Mechanical ventilation in adults with acute respiratory distress
  syndrome. Summary of the experimental evidence for the clinical practice
  guideline}.
\bjournal{Annals of the American Thoracic Society}
\bvolume{14}
\bpages{S261-S270}.
\bnote{PMID: 28985479}.
\bdoi{10.1513/AnnalsATS.201704-345OT}
\end{barticle}
\endbibitem

\bibitem[\protect\citeauthoryear{Deng et~al.}{2021}]{Deng2021}
\begin{barticle}[author]
\bauthor{\bsnm{Deng},~\bfnm{Yuhao}\binits{Y.}},
  \bauthor{\bsnm{Guo},~\bfnm{Yuhang}\binits{Y.}},
  \bauthor{\bsnm{Chang},~\bfnm{Yingjun}\binits{Y.}} \AND
  \bauthor{\bsnm{Zhou},~\bfnm{Xiao-Hua}\binits{X.-H.}}
(\byear{2021}).
\btitle{Identification and estimation of the heterogeneous survivor average
  causal effect in observational studies}.
\bjournal{arXiv preprint arXiv:2109.13623}
\bpages{1-23}.
\end{barticle}
\endbibitem

\bibitem[\protect\citeauthoryear{Ding and Li}{2018}]{Ding2018}
\begin{barticle}[author]
\bauthor{\bsnm{Ding},~\bfnm{Peng}\binits{P.}} \AND
  \bauthor{\bsnm{Li},~\bfnm{Fan}\binits{F.}}
(\byear{2018}).
\btitle{{Causal inference: A missing data perspective}}.
\bjournal{Statistical Science}
\bvolume{33}
\bpages{214-237}.
\bdoi{10.1214/18-STS645}
\end{barticle}
\endbibitem

\bibitem[\protect\citeauthoryear{Ding and Lu}{2017}]{ding2017principal}
\begin{barticle}[author]
\bauthor{\bsnm{Ding},~\bfnm{Peng}\binits{P.}} \AND
  \bauthor{\bsnm{Lu},~\bfnm{Jiannan}\binits{J.}}
(\byear{2017}).
\btitle{Principal stratification analysis using principal scores}.
\bjournal{Journal of the Royal Statistical Society: Series B (Statistical
  Methodology)}
\bvolume{79}
\bpages{757-777}.
\end{barticle}
\endbibitem

\bibitem[\protect\citeauthoryear{Ding et~al.}{2011}]{Ding2011}
\begin{barticle}[author]
\bauthor{\bsnm{Ding},~\bfnm{Peng}\binits{P.}},
  \bauthor{\bsnm{Geng},~\bfnm{Zhi}\binits{Z.}},
  \bauthor{\bsnm{Yan},~\bfnm{Wei}\binits{W.}} \AND
  \bauthor{\bsnm{Zhou},~\bfnm{Xiao-Hua}\binits{X.-H.}}
(\byear{2011}).
\btitle{Identifiability and estimation of causal effects by principal
  stratification with outcomes truncated by death}.
\bjournal{Journal of the American Statistical Association}
\bvolume{106}
\bpages{1578-1591}.
\bdoi{10.1198/jasa.2011.tm10265}
\end{barticle}
\endbibitem

\bibitem[\protect\citeauthoryear{Dorie et~al.}{2019}]{dorie2019automated}
\begin{barticle}[author]
\bauthor{\bsnm{Dorie},~\bfnm{Vincent}\binits{V.}},
  \bauthor{\bsnm{Hill},~\bfnm{Jennifer}\binits{J.}},
  \bauthor{\bsnm{Shalit},~\bfnm{Uri}\binits{U.}},
  \bauthor{\bsnm{Scott},~\bfnm{Marc}\binits{M.}},
  \bauthor{\bsnm{Cervone},~\bfnm{Dan}\binits{D.}} \betal{et~al.}
(\byear{2019}).
\btitle{Automated versus do-it-yourself methods for causal inference: Lessons
  learned from a data analysis competition}.
\bjournal{Statistical Science}
\bvolume{34}
\bpages{43-68}.
\end{barticle}
\endbibitem

\bibitem[\protect\citeauthoryear{Egleston et~al.}{2006}]{Egleston2006}
\begin{barticle}[author]
\bauthor{\bsnm{Egleston},~\bfnm{Brian~L.}\binits{B.~L.}},
  \bauthor{\bsnm{Scharfstein},~\bfnm{Daniel~O.}\binits{D.~O.}},
  \bauthor{\bsnm{Freeman},~\bfnm{Ellen~E.}\binits{E.~E.}} \AND
  \bauthor{\bsnm{West},~\bfnm{Sheila~K.}\binits{S.~K.}}
(\byear{2006}).
\btitle{{Causal inference for non-mortality outcomes in the presence of
  death}}.
\bjournal{Biostatistics}
\bvolume{8}
\bpages{526-545}.
\bdoi{10.1093/biostatistics/kxl027}
\end{barticle}
\endbibitem

\bibitem[\protect\citeauthoryear{Fan et~al.}{2017}]{Fan2017}
\begin{barticle}[author]
\bauthor{\bsnm{Fan},~\bfnm{Eddy}\binits{E.}},
  \bauthor{\bsnm{Del~Sorbo},~\bfnm{Lorenzo}\binits{L.}},
  \bauthor{\bsnm{Goligher},~\bfnm{Ewan~C.}\binits{E.~C.}},
  \bauthor{\bsnm{Hodgson},~\bfnm{Carol~L.}\binits{C.~L.}},
  \bauthor{\bsnm{Munshi},~\bfnm{Laveena}\binits{L.}},
  \bauthor{\bsnm{Walkey},~\bfnm{Allan~J.}\binits{A.~J.}},
  \bauthor{\bsnm{Adhikari},~\bfnm{Neill K.~J.}\binits{N.~K.~J.}},
  \bauthor{\bsnm{Amato},~\bfnm{Marcelo B.~P.}\binits{M.~B.~P.}},
  \bauthor{\bsnm{Branson},~\bfnm{Richard}\binits{R.}},
  \bauthor{\bsnm{Brower},~\bfnm{Roy~G.}\binits{R.~G.}} \betal{et~al.}
(\byear{2017}).
\btitle{An official American thoracic society/European society of intensive
  care medicine/society of critical care medicine clinical practice guideline:
  Mechanical ventilation in adult patients with acute respiratory distress
  syndrome}.
\bjournal{American Journal of Respiratory and Critical Care Medicine}
\bvolume{195}
\bpages{1253-1263}.
\bnote{PMID: 28459336}.
\bdoi{10.1164/rccm.201703-0548ST}
\end{barticle}
\endbibitem

\bibitem[\protect\citeauthoryear{Frangakis and
  Rubin}{2002}]{Frangakis2002biometrics}
\begin{barticle}[author]
\bauthor{\bsnm{Frangakis},~\bfnm{Constantine~E.}\binits{C.~E.}} \AND
  \bauthor{\bsnm{Rubin},~\bfnm{Donald~B.}\binits{D.~B.}}
(\byear{2002}).
\btitle{Principal stratification in causal inference}.
\bjournal{Biometrics}
\bvolume{58}
\bpages{21-29}.
\bdoi{10.1111/j.0006-341X.2002.00021.x}
\end{barticle}
\endbibitem

\bibitem[\protect\citeauthoryear{Frumento et~al.}{2012}]{Frumento2012}
\begin{barticle}[author]
\bauthor{\bsnm{Frumento},~\bfnm{Paolo}\binits{P.}},
  \bauthor{\bsnm{Mealli},~\bfnm{Fabrizia}\binits{F.}},
  \bauthor{\bsnm{Pacini},~\bfnm{Barbara}\binits{B.}} \AND
  \bauthor{\bsnm{Rubin},~\bfnm{Donald~B.}\binits{D.~B.}}
(\byear{2012}).
\btitle{Evaluating the effect of training on wages in the presence of
  noncompliance, nonemployment, and missing outcome data}.
\bjournal{Journal of the American Statistical Association}
\bvolume{107}
\bpages{450-466}.
\bdoi{10.2307/23239583}
\end{barticle}
\endbibitem

\bibitem[\protect\citeauthoryear{Goligher et~al.}{2021}]{goligher2021effect}
\begin{barticle}[author]
\bauthor{\bsnm{Goligher},~\bfnm{Ewan~C}\binits{E.~C.}},
  \bauthor{\bsnm{Costa},~\bfnm{Eduardo~LV}\binits{E.~L.}},
  \bauthor{\bsnm{Yarnell},~\bfnm{Christopher~J}\binits{C.~J.}},
  \bauthor{\bsnm{Brochard},~\bfnm{Laurent~J}\binits{L.~J.}},
  \bauthor{\bsnm{Stewart},~\bfnm{Thomas~E}\binits{T.~E.}},
  \bauthor{\bsnm{Tomlinson},~\bfnm{George}\binits{G.}},
  \bauthor{\bsnm{Brower},~\bfnm{Roy~G}\binits{R.~G.}},
  \bauthor{\bsnm{Slutsky},~\bfnm{Arthur~S}\binits{A.~S.}} \AND
  \bauthor{\bsnm{Amato},~\bfnm{Marcelo~PB}\binits{M.~P.}}
(\byear{2021}).
\btitle{Effect of lowering Vt on mortality in acute respiratory distress
  syndrome varies with respiratory system elastance}.
\bjournal{American Journal of Respiratory and Critical Care Medicine}
\bvolume{203}
\bpages{1378-1385}.
\end{barticle}
\endbibitem

\bibitem[\protect\citeauthoryear{Hahn, Murray and
  Manolopoulou}{2016}]{Hahn2016}
\begin{barticle}[author]
\bauthor{\bsnm{Hahn},~\bfnm{P.~Richard}\binits{P.~R.}},
  \bauthor{\bsnm{Murray},~\bfnm{Jared~S.}\binits{J.~S.}} \AND
  \bauthor{\bsnm{Manolopoulou},~\bfnm{Ioanna}\binits{I.}}
(\byear{2016}).
\btitle{{A Bayesian partial identification approach to inferring the prevalence
  of accounting misconduct}}.
\bjournal{Journal of the American Statistical Association}
\bvolume{111}
\bpages{14-26}.
\bdoi{10.1080/01621459.2015.1084307}
\end{barticle}
\endbibitem

\bibitem[\protect\citeauthoryear{Hahn, Murray and Carvalho}{2020}]{Hahn2020}
\begin{barticle}[author]
\bauthor{\bsnm{Hahn},~\bfnm{P.~Richard}\binits{P.~R.}},
  \bauthor{\bsnm{Murray},~\bfnm{Jared~S.}\binits{J.~S.}} \AND
  \bauthor{\bsnm{Carvalho},~\bfnm{Carlos~M.}\binits{C.~M.}}
(\byear{2020}).
\btitle{Bayesian regression tree models for causal inference: Regularization,
  confounding, and heterogeneous effects (with discussion)}.
\bjournal{Bayesian Analysis}
\bvolume{15}
\bpages{965-1056}.
\bdoi{10.1214/19-BA1195}
\end{barticle}
\endbibitem

\bibitem[\protect\citeauthoryear{Harhay et~al.}{2014}]{harhay2014outcomes}
\begin{barticle}[author]
\bauthor{\bsnm{Harhay},~\bfnm{Michael~O}\binits{M.~O.}},
  \bauthor{\bsnm{Wagner},~\bfnm{Jason}\binits{J.}},
  \bauthor{\bsnm{Ratcliffe},~\bfnm{Sarah~J}\binits{S.~J.}},
  \bauthor{\bsnm{Bronheim},~\bfnm{Rachel~S}\binits{R.~S.}},
  \bauthor{\bsnm{Gopal},~\bfnm{Anand}\binits{A.}},
  \bauthor{\bsnm{Green},~\bfnm{Sydney}\binits{S.}},
  \bauthor{\bsnm{Cooney},~\bfnm{Elizabeth}\binits{E.}},
  \bauthor{\bsnm{Mikkelsen},~\bfnm{Mark~E}\binits{M.~E.}},
  \bauthor{\bsnm{Kerlin},~\bfnm{Meeta~Prasad}\binits{M.~P.}},
  \bauthor{\bsnm{Small},~\bfnm{Dylan~S}\binits{D.~S.}} \betal{et~al.}
(\byear{2014}).
\btitle{Outcomes and statistical power in adult critical care randomized
  trials}.
\bjournal{American Journal of Respiratory and Critical Care Medicine}
\bvolume{189}
\bpages{1469-1478}.
\end{barticle}
\endbibitem

\bibitem[\protect\citeauthoryear{Harhay et~al.}{2019}]{Harhay2019}
\begin{barticle}[author]
\bauthor{\bsnm{Harhay},~\bfnm{Michael~O.}\binits{M.~O.}},
  \bauthor{\bsnm{Ratcliffe},~\bfnm{Sarah~J.}\binits{S.~J.}},
  \bauthor{\bsnm{Small},~\bfnm{Dylan~S.}\binits{D.~S.}},
  \bauthor{\bsnm{Suttner},~\bfnm{Leah~H.}\binits{L.~H.}},
  \bauthor{\bsnm{Crowther},~\bfnm{Michael~J.}\binits{M.~J.}} \AND
  \bauthor{\bsnm{Halpern},~\bfnm{Scott~D.}\binits{S.~D.}}
(\byear{2019}).
\btitle{Measuring and analyzing length of stay in critical care trials}.
\bjournal{Medical Care}
\bvolume{57}
\bpages{e53-e59}.
\bnote{PMID: 30664613}.
\bdoi{10.1097/MLR.0000000000001059}
\end{barticle}
\endbibitem

\bibitem[\protect\citeauthoryear{Hayden, Pauler and
  Schoenfeld}{2005}]{Hayden2005}
\begin{barticle}[author]
\bauthor{\bsnm{Hayden},~\bfnm{Douglas}\binits{D.}},
  \bauthor{\bsnm{Pauler},~\bfnm{Donna~K.}\binits{D.~K.}} \AND
  \bauthor{\bsnm{Schoenfeld},~\bfnm{David}\binits{D.}}
(\byear{2005}).
\btitle{An estimator for treatment comparisons among survivors in randomized
  trials}.
\bjournal{Biometrics}
\bvolume{61}
\bpages{305-310}.
\bdoi{https://doi.org/10.1111/j.0006-341X.2005.030227.x}
\end{barticle}
\endbibitem

\bibitem[\protect\citeauthoryear{{Helmholz Jr.}}{1979}]{Helmholz1979}
\begin{barticle}[author]
\bauthor{\bsnm{{Helmholz Jr. }},~\bfnm{H.~Frederic}\binits{H.~F.}}
(\byear{1979}).
\btitle{The abbreviated alveolar air equation}.
\bjournal{Chest}
\bvolume{75}
\bpages{748}.
\bdoi{doi: 10.1378/chest.75.6.748}
\end{barticle}
\endbibitem

\bibitem[\protect\citeauthoryear{Henderson et~al.}{2018}]{Henderson2018}
\begin{barticle}[author]
\bauthor{\bsnm{Henderson},~\bfnm{Nicholas~C}\binits{N.~C.}},
  \bauthor{\bsnm{Louis},~\bfnm{Thomas~A}\binits{T.~A.}},
  \bauthor{\bsnm{Rosner},~\bfnm{Gary~L}\binits{G.~L.}} \AND
  \bauthor{\bsnm{Varadhan},~\bfnm{Ravi}\binits{R.}}
(\byear{2018}).
\btitle{Individualized treatment effects with censored data via fully
  nonparametric Bayesian accelerated failure time models}.
\bjournal{Biostatistics}
\bvolume{21}
\bpages{50-68}.
\bdoi{10.1093/biostatistics/kxy028}
\end{barticle}
\endbibitem

\bibitem[\protect\citeauthoryear{Hill}{2011}]{Hill2011}
\begin{barticle}[author]
\bauthor{\bsnm{Hill},~\bfnm{Jennifer~L.}\binits{J.~L.}}
(\byear{2011}).
\btitle{Bayesian nonparametric modeling for causal inference}.
\bjournal{Journal of Computational and Graphical Statistics}
\bvolume{20}
\bpages{217-240}.
\bdoi{10.1198/jcgs.2010.08162}
\end{barticle}
\endbibitem

\bibitem[\protect\citeauthoryear{Hirano et~al.}{2000}]{hirano2000assessing}
\begin{barticle}[author]
\bauthor{\bsnm{Hirano},~\bfnm{Keisuke}\binits{K.}},
  \bauthor{\bsnm{Imbens},~\bfnm{Guido~W}\binits{G.~W.}},
  \bauthor{\bsnm{Rubin},~\bfnm{Donald~B}\binits{D.~B.}} \AND
  \bauthor{\bsnm{Zhou},~\bfnm{Xiao-Hua}\binits{X.-H.}}
(\byear{2000}).
\btitle{Assessing the effect of an influenza vaccine in an encouragement
  design}.
\bjournal{Biostatistics}
\bvolume{1}
\bpages{69-88}.
\end{barticle}
\endbibitem

\bibitem[\protect\citeauthoryear{Hu, Ji and Li}{2021}]{Hu2021}
\begin{barticle}[author]
\bauthor{\bsnm{Hu},~\bfnm{Liangyuan}\binits{L.}},
  \bauthor{\bsnm{Ji},~\bfnm{Jiayi}\binits{J.}} \AND
  \bauthor{\bsnm{Li},~\bfnm{Fan}\binits{F.}}
(\byear{2021}).
\btitle{Estimating heterogeneous survival treatment effect in observational
  data using machine learning}.
\bjournal{Statistics in Medicine}
\bvolume{40}
\bpages{4691-4713}.
\bdoi{10.1002/sim.9090}
\end{barticle}
\endbibitem

\bibitem[\protect\citeauthoryear{Imai}{2008}]{imai2008sharp}
\begin{barticle}[author]
\bauthor{\bsnm{Imai},~\bfnm{Kosuke}\binits{K.}}
(\byear{2008}).
\btitle{Sharp bounds on the causal effects in randomized experiments with
  ``truncation-by-death"}.
\bjournal{Statistics \& Probability Letters}
\bvolume{78}
\bpages{144-149}.
\end{barticle}
\endbibitem

\bibitem[\protect\citeauthoryear{Kadane}{1975}]{Kadane1975}
\begin{bincollection}[author]
\bauthor{\bsnm{Kadane},~\bfnm{Joseph~Born}\binits{J.~B.}}
(\byear{1975}).
\btitle{{The role of identification in Bayesian theory}}.
In \bbooktitle{Studies in Bayesian Econometrics and Statistics}
(\beditor{\bfnm{S~E}\binits{S.~E.}~\bsnm{Fienberg}} \AND
  \beditor{\bfnm{A}\binits{A.}~\bsnm{Zellner}}, eds.)
\bchapter{5.2},
\bpages{175-191}.
\bpublisher{Amsterdam: North-Holland}.
\end{bincollection}
\endbibitem

\bibitem[\protect\citeauthoryear{Kim et~al.}{2017}]{kim2017framework}
\begin{barticle}[author]
\bauthor{\bsnm{Kim},~\bfnm{Chanmin}\binits{C.}},
  \bauthor{\bsnm{Daniels},~\bfnm{Michael~J}\binits{M.~J.}},
  \bauthor{\bsnm{Marcus},~\bfnm{Bess~H}\binits{B.~H.}} \AND
  \bauthor{\bsnm{Roy},~\bfnm{Jason~A}\binits{J.~A.}}
(\byear{2017}).
\btitle{A framework for Bayesian nonparametric inference for causal effects of
  mediation}.
\bjournal{Biometrics}
\bvolume{73}
\bpages{401-409}.
\end{barticle}
\endbibitem

\bibitem[\protect\citeauthoryear{Kim et~al.}{2019}]{kim2019bayesian}
\begin{barticle}[author]
\bauthor{\bsnm{Kim},~\bfnm{Chanmin}\binits{C.}},
  \bauthor{\bsnm{Daniels},~\bfnm{Michael~J}\binits{M.~J.}},
  \bauthor{\bsnm{Hogan},~\bfnm{Joseph~W}\binits{J.~W.}},
  \bauthor{\bsnm{Choirat},~\bfnm{Christine}\binits{C.}} \AND
  \bauthor{\bsnm{Zigler},~\bfnm{Corwin~M}\binits{C.~M.}}
(\byear{2019}).
\btitle{Bayesian methods for multiple mediators: Relating principal
  stratification and causal mediation in the analysis of power plant emission
  controls}.
\bjournal{The Annals of Applied Statistics}
\bvolume{13}
\bpages{1927}.
\end{barticle}
\endbibitem

\bibitem[\protect\citeauthoryear{Li and Li}{2019}]{Li2019}
\begin{barticle}[author]
\bauthor{\bsnm{Li},~\bfnm{Fan}\binits{F.}} \AND
  \bauthor{\bsnm{Li},~\bfnm{Fan}\binits{F.}}
(\byear{2019}).
\btitle{{Propensity score weighting for causal inference with multiple
  treatments}}.
\bjournal{The Annals of Applied Statistics}
\bvolume{13}
\bpages{2389 - 2415}.
\bdoi{10.1214/19-AOAS1282}
\end{barticle}
\endbibitem

\bibitem[\protect\citeauthoryear{Long and Hudgens}{2013}]{long2013sharpening}
\begin{barticle}[author]
\bauthor{\bsnm{Long},~\bfnm{Dustin~M}\binits{D.~M.}} \AND
  \bauthor{\bsnm{Hudgens},~\bfnm{Michael~G}\binits{M.~G.}}
(\byear{2013}).
\btitle{Sharpening bounds on principal effects with covariates}.
\bjournal{Biometrics}
\bvolume{69}
\bpages{812-819}.
\end{barticle}
\endbibitem

\bibitem[\protect\citeauthoryear{Mattei, Li and Mealli}{2013}]{Mattei2013}
\begin{barticle}[author]
\bauthor{\bsnm{Mattei},~\bfnm{Alessandra}\binits{A.}},
  \bauthor{\bsnm{Li},~\bfnm{Fan}\binits{F.}} \AND
  \bauthor{\bsnm{Mealli},~\bfnm{Fabrizia}\binits{F.}}
(\byear{2013}).
\btitle{Exploiting multiple outcomes in Bayesian principal stratification
  analysis with application to the evaluation of a job training program}.
\bjournal{The Annals of Applied Statistics}
\bvolume{7}
\bpages{2336-2360}.
\bdoi{10.1214/13-AOAS674}
\end{barticle}
\endbibitem

\bibitem[\protect\citeauthoryear{Mattei and Mealli}{2007}]{Mattei2007}
\begin{barticle}[author]
\bauthor{\bsnm{Mattei},~\bfnm{A.}\binits{A.}} \AND
  \bauthor{\bsnm{Mealli},~\bfnm{F.}\binits{F.}}
(\byear{2007}).
\btitle{Application of the principal stratification approach to the Faenza
  randomized experiment on breast self-examination}.
\bjournal{Biometrics}
\bvolume{63}
\bpages{437-446}.
\bdoi{https://doi.org/10.1111/j.1541-0420.2006.00684.x}
\end{barticle}
\endbibitem

\bibitem[\protect\citeauthoryear{Matthay, McAuley and
  Ware}{2017}]{matthay2017clinical}
\begin{barticle}[author]
\bauthor{\bsnm{Matthay},~\bfnm{Michael~A}\binits{M.~A.}},
  \bauthor{\bsnm{McAuley},~\bfnm{Daniel~F}\binits{D.~F.}} \AND
  \bauthor{\bsnm{Ware},~\bfnm{Lorraine~B}\binits{L.~B.}}
(\byear{2017}).
\btitle{Clinical trials in acute respiratory distress syndrome: challenges and
  opportunities}.
\bjournal{The Lancet Respiratory Medicine}
\bvolume{5}
\bpages{524-534}.
\end{barticle}
\endbibitem

\bibitem[\protect\citeauthoryear{McCaffrey
  et~al.}{2013}]{mccaffrey2013tutorial}
\begin{barticle}[author]
\bauthor{\bsnm{McCaffrey},~\bfnm{Daniel~F}\binits{D.~F.}},
  \bauthor{\bsnm{Griffin},~\bfnm{Beth~Ann}\binits{B.~A.}},
  \bauthor{\bsnm{Almirall},~\bfnm{Daniel}\binits{D.}},
  \bauthor{\bsnm{Slaughter},~\bfnm{Mary~Ellen}\binits{M.~E.}},
  \bauthor{\bsnm{Ramchand},~\bfnm{Rajeev}\binits{R.}} \AND
  \bauthor{\bsnm{Burgette},~\bfnm{Lane~F}\binits{L.~F.}}
(\byear{2013}).
\btitle{A tutorial on propensity score estimation for multiple treatments using
  generalized boosted models}.
\bjournal{Statistics in Medicine}
\bvolume{32}
\bpages{3388-3414}.
\end{barticle}
\endbibitem

\bibitem[\protect\citeauthoryear{McNicholas et~al.}{2019}]{McNicholas2019}
\begin{barticle}[author]
\bauthor{\bsnm{McNicholas},~\bfnm{Bairbre~A.}\binits{B.~A.}},
  \bauthor{\bsnm{Madotto},~\bfnm{Fabiana}\binits{F.}},
  \bauthor{\bsnm{Pham},~\bfnm{T{\`a}i}\binits{T.}},
  \bauthor{\bsnm{Rezoagli},~\bfnm{Emanuele}\binits{E.}},
  \bauthor{\bsnm{Masterson},~\bfnm{Claire~H.}\binits{C.~H.}},
  \bauthor{\bsnm{Horie},~\bfnm{Shahd}\binits{S.}},
  \bauthor{\bsnm{Bellani},~\bfnm{Giacomo}\binits{G.}},
  \bauthor{\bsnm{Brochard},~\bfnm{Laurent}\binits{L.}} \AND
  \bauthor{\bsnm{Laffey},~\bfnm{John~G.}\binits{J.~G.}}
(\byear{2019}).
\btitle{Demographics, management and outcome of women and men with Acute
  Respiratory Distress Syndrome in the LUNG SAFE prospective cohort study}.
\bjournal{European Respiratory Journal}
\bvolume{54}.
\bnote{PMID: 31346004}.
\bdoi{10.1183/13993003.00609-2019}
\end{barticle}
\endbibitem

\bibitem[\protect\citeauthoryear{Nevo and Gorfine}{2021}]{Nevo2021}
\begin{barticle}[author]
\bauthor{\bsnm{Nevo},~\bfnm{Daniel}\binits{D.}} \AND
  \bauthor{\bsnm{Gorfine},~\bfnm{Malka}\binits{M.}}
(\byear{2021}).
\btitle{{Causal inference for semi-competing risks data}}.
\bjournal{Biostatistics}
\bvolume{23}
\bpages{1115-1132}.
\bdoi{10.1093/biostatistics/kxab049}
\end{barticle}
\endbibitem

\bibitem[\protect\citeauthoryear{Papakostas et~al.}{2022}]{Papakostas2021}
\begin{barticle}[author]
\bauthor{\bsnm{Papakostas},~\bfnm{Demetrios}\binits{D.}},
  \bauthor{\bsnm{Hahn},~\bfnm{P.~Richard}\binits{P.~R.}},
  \bauthor{\bsnm{Murray},~\bfnm{Jared~S}\binits{J.~S.}},
  \bauthor{\bsnm{Zhou},~\bfnm{Frank}\binits{F.}} \AND
  \bauthor{\bsnm{Gerakos},~\bfnm{Joseph}\binits{J.}}
(\byear{2022}).
\btitle{{Do forecasts of bankruptcy cause bankruptcy? A machine learning
  sensitivity analysis}}.
\bjournal{The Annals of Applied Statistics}
\bpages{1-31}.
\end{barticle}
\endbibitem

\bibitem[\protect\citeauthoryear{Roy, Lum and Daniels}{2017}]{roy2017bayesian}
\begin{barticle}[author]
\bauthor{\bsnm{Roy},~\bfnm{Jason}\binits{J.}},
  \bauthor{\bsnm{Lum},~\bfnm{Kirsten~J}\binits{K.~J.}} \AND
  \bauthor{\bsnm{Daniels},~\bfnm{Michael~J}\binits{M.~J.}}
(\byear{2017}).
\btitle{A Bayesian nonparametric approach to marginal structural models for
  point treatments and a continuous or survival outcome}.
\bjournal{Biostatistics}
\bvolume{18}
\bpages{32-47}.
\end{barticle}
\endbibitem

\bibitem[\protect\citeauthoryear{Shen et~al.}{2019}]{Shen2019}
\begin{barticle}[author]
\bauthor{\bsnm{Shen},~\bfnm{Yanfei}\binits{Y.}},
  \bauthor{\bsnm{Cai},~\bfnm{Guolong}\binits{G.}},
  \bauthor{\bsnm{Gong},~\bfnm{Shijin}\binits{S.}},
  \bauthor{\bsnm{Dong},~\bfnm{Lei}\binits{L.}},
  \bauthor{\bsnm{Yan},~\bfnm{Jing}\binits{J.}} \AND
  \bauthor{\bsnm{Cai},~\bfnm{Wanru}\binits{W.}}
(\byear{2019}).
\btitle{Interaction between low tidal volume ventilation strategy and severity
  of acute respiratory distress syndrome: A retrospective cohort study}.
\bjournal{Critical Care}
\bvolume{23}
\bpages{254}.
\bdoi{10.1186/s13054-019-2530-6}
\end{barticle}
\endbibitem

\bibitem[\protect\citeauthoryear{Tan and Roy}{2019}]{tan2019bayesian}
\begin{barticle}[author]
\bauthor{\bsnm{Tan},~\bfnm{Yaoyuan~Vincent}\binits{Y.~V.}} \AND
  \bauthor{\bsnm{Roy},~\bfnm{Jason}\binits{J.}}
(\byear{2019}).
\btitle{Bayesian additive regression trees and the general BART model}.
\bjournal{Statistics in Medicine}
\bvolume{38}
\bpages{5048-5069}.
\end{barticle}
\endbibitem

\bibitem[\protect\citeauthoryear{Tonelli et~al.}{2014}]{Tonelli2014}
\begin{barticle}[author]
\bauthor{\bsnm{Tonelli},~\bfnm{Adriano~R.}\binits{A.~R.}},
  \bauthor{\bsnm{Zein},~\bfnm{Joe}\binits{J.}},
  \bauthor{\bsnm{Adams},~\bfnm{Jacob}\binits{J.}} \AND
  \bauthor{\bsnm{Ioannidis},~\bfnm{John P.~A.}\binits{J.~P.~A.}}
(\byear{2014}).
\btitle{Effects of interventions on survival in acute respiratory distress
  syndrome: An umbrella review of 159 published randomized trials and 29
  meta-analyses}.
\bjournal{Intensive Care Medicine}
\bvolume{40}
\bpages{769-787}.
\bdoi{10.1007/s00134-014-3272-1}
\end{barticle}
\endbibitem

\bibitem[\protect\citeauthoryear{Wang, Zhou and
  Richardson}{2017}]{wang2017identification}
\begin{barticle}[author]
\bauthor{\bsnm{Wang},~\bfnm{Linbo}\binits{L.}},
  \bauthor{\bsnm{Zhou},~\bfnm{Xiao-Hua}\binits{X.-H.}} \AND
  \bauthor{\bsnm{Richardson},~\bfnm{Thomas~S}\binits{T.~S.}}
(\byear{2017}).
\btitle{Identification and estimation of causal effects with outcomes truncated
  by death}.
\bjournal{Biometrika}
\bvolume{104}
\bpages{597-612}.
\end{barticle}
\endbibitem

\bibitem[\protect\citeauthoryear{Wendling et~al.}{2018}]{wendling2018comparing}
\begin{barticle}[author]
\bauthor{\bsnm{Wendling},~\bfnm{Thierry}\binits{T.}},
  \bauthor{\bsnm{Jung},~\bfnm{Kenneth}\binits{K.}},
  \bauthor{\bsnm{Callahan},~\bfnm{Alison}\binits{A.}},
  \bauthor{\bsnm{Schuler},~\bfnm{Alejandro}\binits{A.}},
  \bauthor{\bsnm{Shah},~\bfnm{Nigam~H}\binits{N.~H.}} \AND
  \bauthor{\bsnm{Gallego},~\bfnm{Blanco}\binits{B.}}
(\byear{2018}).
\btitle{Comparing methods for estimation of heterogeneous treatment effects
  using observational data from health care databases}.
\bjournal{Statistics in Medicine}
\bvolume{37}
\bpages{3309-3324}.
\end{barticle}
\endbibitem

\bibitem[\protect\citeauthoryear{Woody, Carvalho and Murray}{2021}]{Woody2021}
\begin{barticle}[author]
\bauthor{\bsnm{Woody},~\bfnm{Spencer}\binits{S.}},
  \bauthor{\bsnm{Carvalho},~\bfnm{Carlos~M.}\binits{C.~M.}} \AND
  \bauthor{\bsnm{Murray},~\bfnm{Jared~S.}\binits{J.~S.}}
(\byear{2021}).
\btitle{Model Interpretation Through Lower-Dimensional Posterior
  Summarization}.
\bjournal{Journal of Computational and Graphical Statistics}
\bvolume{30}
\bpages{144-161}.
\bdoi{10.1080/10618600.2020.1796684}
\end{barticle}
\endbibitem

\bibitem[\protect\citeauthoryear{Xu et~al.}{2016}]{xu2016bayesian}
\begin{barticle}[author]
\bauthor{\bsnm{Xu},~\bfnm{Yanxun}\binits{Y.}},
  \bauthor{\bsnm{M{\"u}ller},~\bfnm{Peter}\binits{P.}},
  \bauthor{\bsnm{Wahed},~\bfnm{Abdus~S}\binits{A.~S.}} \AND
  \bauthor{\bsnm{Thall},~\bfnm{Peter~F}\binits{P.~F.}}
(\byear{2016}).
\btitle{Bayesian nonparametric estimation for dynamic treatment regimes with
  sequential transition times}.
\bjournal{Journal of the American Statistical Association}
\bvolume{111}
\bpages{921-950}.
\end{barticle}
\endbibitem

\bibitem[\protect\citeauthoryear{Xu et~al.}{2020}]{Xu2022}
\begin{barticle}[author]
\bauthor{\bsnm{Xu},~\bfnm{Yanxun}\binits{Y.}},
  \bauthor{\bsnm{Scharfstein},~\bfnm{Daniel}\binits{D.}},
  \bauthor{\bsnm{M\"uller},~\bfnm{Peter}\binits{P.}} \AND
  \bauthor{\bsnm{Daniels},~\bfnm{Michael}\binits{M.}}
(\byear{2020}).
\btitle{{A Bayesian nonparametric approach for evaluating the causal effect of
  treatment in randomized trials with semi-competing risks}}.
\bjournal{Biostatistics}
\bvolume{23}
\bpages{34-49}.
\bdoi{10.1093/biostatistics/kxaa008}
\end{barticle}
\endbibitem

\bibitem[\protect\citeauthoryear{Yang and Small}{2016}]{Yang2016}
\begin{barticle}[author]
\bauthor{\bsnm{Yang},~\bfnm{Fan}\binits{F.}} \AND
  \bauthor{\bsnm{Small},~\bfnm{Dylan~S.}\binits{D.~S.}}
(\byear{2016}).
\btitle{Using post-outcome measurement information in censoring-by-death
  problems}.
\bjournal{Journal of the Royal Statistical Society: Series B (Statistical
  Methodology)}
\bvolume{78}
\bpages{299-318}.
\bdoi{10.1111/rssb.12113}
\end{barticle}
\endbibitem

\bibitem[\protect\citeauthoryear{Zhang and Rubin}{2003}]{ZhangRubin2003}
\begin{barticle}[author]
\bauthor{\bsnm{Zhang},~\bfnm{Junni~L.}\binits{J.~L.}} \AND
  \bauthor{\bsnm{Rubin},~\bfnm{Donald~B.}\binits{D.~B.}}
(\byear{2003}).
\btitle{Estimation of causal effects via principal stratification when some
  outcomes are truncated by ``death''}.
\bjournal{Journal of Educational and Behavioral Statistics}
\bvolume{28}
\bpages{353-368}.
\bdoi{10.3102/10769986028004353}
\end{barticle}
\endbibitem

\bibitem[\protect\citeauthoryear{Zhang, Rubin and Mealli}{2009}]{Zhang2009jasa}
\begin{barticle}[author]
\bauthor{\bsnm{Zhang},~\bfnm{Junni~L.}\binits{J.~L.}},
  \bauthor{\bsnm{Rubin},~\bfnm{Donald~B.}\binits{D.~B.}} \AND
  \bauthor{\bsnm{Mealli},~\bfnm{Fabrizia}\binits{F.}}
(\byear{2009}).
\btitle{Likelihood-based analysis of causal effects of job-training programs
  using principal stratification}.
\bjournal{Journal of the American Statistical Association}
\bvolume{104}
\bpages{166-176}.
\bdoi{10.1198/jasa.2009.0012}
\end{barticle}
\endbibitem

\end{thebibliography}

\end{document}